\documentclass[12pt]{spieman}  
\usepackage{amsmath,amsfonts,amssymb}
\usepackage{graphicx}
\usepackage{setspace}
\usepackage{tocloft}
\usepackage{siunitx}
\usepackage{array}
\usepackage{float} 
\usepackage[T1]{fontenc}
\usepackage[utf8]{inputenc}
\newcolumntype{L}[1]{>{\raggedright\arraybackslash}p{#1}}

\title{Stimulated interactions of low-energy free-electrons with light}
\author[a,+]{Fatemeh Chahshouri}
\author[b]{Sven Ebel}
\author[a]{Mitja Funk}
\author[a,c,*]{Nahid Talebi}

\affil[a]{Institute of Experimental and Applied Physics, Kiel University, 24098 Kiel, Germany}
\affil[b]{POLIMA---Center for Polariton-driven Light--Matter Interactions, University of Southern Denmark, 5230 Odense M, Denmark} 
\affil[c]{Kiel Nano, Surface and Interface Science KiNSIS, Kiel University, 24118 Kiel, Germany
}%

\cftpagenumbersoff{figure}
\cftpagenumbersoff{table} 
\begin{document} 
\maketitle

\begin{abstract}

Free-electron interactions with light and matter have long served as a cornerstone for exploring the quantum and ultrafast dynamics of material excitation. In recent years, this paradigm has evolved from a classical description of radiation and acceleration toward a fully quantum framework, transforming our understanding of light–matter interactions at the single-electron level. These advances have opened new opportunities in high-resolution imaging, ultrafast spectroscopy/interferometry, and the coherent shaping of electron wavepackets. \par
This review surveys stimulated interactions between slow electrons and light, encompassing free-space and near-field mediated mechanisms. We discuss how free-space optical fields coherently modulate electron momentum and energy, and how near-field coupling in nanophotonic and plasmonic structures enables strong, phase-matched, efficient momentum exchange with the electron wavepacket. We further describe electron recoil, which is significant in the slow-electron regime, and temporal and spatial wavepacket shaping that enhances coupling efficiency and extends access to quantum-coherent regimes. Building on these foundations, we outline emerging frameworks including hybrid optical–electrostatic modulation, ponderomotive laser-based aberration correction, and optical electron interferometry. By unifying these developments, stimulated electron–light interactions provide a versatile route to precise beam control, quantum-state engineering, and tailored light–matter coupling, with implications for ultrafast spectroscopy, nanoscale metrology, attosecond pulse generation, electron–photon entanglement, and the creation of nonclassical states of light.

\end{abstract}


{\noindent \footnotesize\textbf{+} Fatemeh Chahshouri,  \linkable{Chahshouri@physik.uni-kiel.de} }

{\noindent \footnotesize\textbf{*}Nahid Talebi,  \linkable{Talebi@physik.uni-kiel.de} }

\renewcommand{\contentsname}{Table of Contents}
\tableofcontents


\section{\label{sec:level1}Introduction}
The interaction between free electrons and light has been a fundamental concept in physics for more than a century, from the earliest scattering experiments \cite{THOMSON1927DiffractionFilm,davisson_diffraction_1927} to the development of electron microscopy and diffraction \cite{Knoll1932DasElektronenmikroskop,Vanacore2016Four-dimensionalBiology}. In its traditional formulation, this interaction was described within a classical framework, where electrons were treated as point-like particles deflected by electromagnetic fields \cite{talebi_schrodinger_2016,Shiloh2022MiniatureControl,Pan2023WeakInteractions}. While this approach accounted for many early observations, it overlooked the intrinsic wave nature of electrons \cite{barwick_photon-induced_2009,Dahan2020ResonantWavefunction,feist_quantum_2015}, for describing phenomena such as quantum-path interferences, quantum
nonlinearity, and decoherence. In recent years, innovations in electron microscopy have further transformed nanoscience by enabling atomic-scale insights into biological, chemical, and semiconductor materials \cite{polman_electron-beam_2019}. With all this progress, the quantum-coherent interaction of free electrons and photons is currently evolving into a booming research field. It started with the advances of photon-induced near-field electron microscopy (PINEM) in transmission electron microscopes (TEMs) by integrating a femtosecond laser pulse with an electron microscope two decades ago\cite{barwick_4d_2008,barwick_photon-induced_2009}. Since then, electron microscopes have been developed not only as devices for investigating the ultrafast dynamics of material excitations such as localized plasmons\cite{Yurtsever2012SubparticleMicroscopy}, and phonon polaritons \cite{Kurman2021SpatiotemporalElectrons}, but also into an apparatus to test the frontiers of quantum science, such as strong-coupling effects\cite{Wang2020CoherentCavity}, electron-photon entanglement \cite{kfir_entanglements_2019}, ultrafast charge oscillations \cite{park_photon-induced_2010,Hergert2021ProbingMicroscopy}, and nonequilibrium optical excitations \cite{Kurman2021SpatiotemporalElectrons, Madan2023ChargeDynamics}.  These advances have opened new ways for probing plasmon resonances \cite{Wang2020CoherentCavity,talebi_strong_2020,Das2019StimulatedGun}, exciton dynamics \cite{Taleb2023Phase-lockedLaser}, and phonon behavior \cite{Kurman2021SpatiotemporalElectrons}, thus driving breakthroughs in electron holography \cite{Vogelsang2018ObservingMicroscopy, Madan2019HolographicInterference}, phase retrieval \cite{priebe_attosecond_2017}, attosecond pulse trains \cite{priebe_attosecond_2017,kozak_ponderomotive_2018}, and wave packet shaping \cite{chahshouri_numerical_2023,chahshouri_tailoring_2023,DiGiulio2020Free-electronLight}. \par
To date, most investigations of stimulated electron–photon interactions have employed fast, relativistic electrons in the 70–300 keV range. These high-energy probes have been essential for mapping optical near-fields and quantifying light–matter coupling strengths. Extending this research to slower, low-energy electrons, however, opens a distinct regime in which recoil effects become significant and interaction times are substantially prolonged. Building on this conceptual and instrumental foundation, recent experiments have begun to explore these effects using much less energetic electrons in scanning electron microscopes (SEMs) \cite{shiloh_quantum-coherent_2022,talebi_strong_2020} and point-projection microscopes (PPMs) \cite{Hergert2021ProbingMicroscopy,Hergert2021StrongNanostructures}. In this slow-electron regime (below 30 keV), the reduced electron velocity enhances phase synchronization and extends the interaction time so that even moderate optical fields can induce a pronounced energy–momentum exchange. The recoil effect becomes substantial \cite{Synanidis2024QuantumLight,RoquesCarmes2023,Talebi2018Electron-lightInterferometry}, giving rise to asymmetric sidebands, transverse deflections, and richer quantum-coherent pathways. Taken together, these features make slow electrons ideally suited for investigating strong coupling, recoil-sensitive dynamics\cite{talebi_near-field-mediated_2019,RoquesCarmes2023,Talebi2018Electron-lightInterferometry}, and programmable wavefunction shaping in experimentally accessible and compact systems. \par
Understanding the fundamentals of these interactions requires a fully quantum-mechanical framework, achieved by combining time-dependent Schrödinger-equation solvers with Maxwell-based electromagnetic field models \cite{Talebi2018Electron-lightInterferometry}. Such approaches reveal a rich landscape of new physics, from strong-coupling signatures\cite{talebi_strong_2020} reminiscent of cavity quantum electrodynamics to quantum-path interactions\cite{talebi_interference_2019}.
Ultrafast energy gating, in addition, enables femtosecond-resolved imaging of transient phenomena such as carrier transport, phonon generation, and plasmon oscillations \cite{Madan2023ChargeDynamics,Wang2020CoherentCavity}. 
Drawing on concepts from quantum optics \cite{feist_quantum_2015,Bucher2024CoherentlyInterferometer,Wang2020CoherentCavity}, researchers have developed methods to manipulate electron wave functions with these confined fields.\cite{baum_attosecond_2007,vanacore_attosecond_2018} 
The coherence of an electron beam is characterized by two orthogonal components: longitudinal (temporal) and transverse (spatial) coherence. Longitudinal coherence describes the phase relationship along the direction of propagation and is inversely proportional to the energy spread ($\Delta E$) of the electrons, defining the temporal coherence of the wavepacket. In contrast, transverse coherence refers to the phase relationship between different points across the beam's wavefront, relating the spatial coherence directly to the source size and beam divergence \cite{Ehberger2015HighlyTip}. These parameters are fundamental in determining spatial resolution and interference contrast in electron-based interferometry and holography experiments \cite{Feist2017UTEMFieldEmitter,Spence2013HREM}. By engineering both the longitudinal phase, to enable temporal compression, spectral sideband engineering, or attosecond pulse trains\cite{baum_attosecond_2007}, and the transverse phase\cite{Kozak2016TransverseNear-fields,Kozak2021ElectronScattering}, to correct aberrations or generate structured beams such as vortex states \cite{uchida_generation_2010,McMorran2011ElectronMomentum}, it is possible to mitigate the effects of diffraction and space-charge broadening while enhancing spatial coherence.\par 

Coherent light fields can imprint longitudinal\cite{feist_quantum_2015,henke_integrated_2021} and transverse \cite{vanacore_attosecond_2018,tsesses_tunable_2023,madan_ultrafast_2022,chirita_mihaila_transverse_2022} phase modulations on electron wavepackets. This can occur in free space via the vector potential\cite{talebi_near-field-mediated_2019,freimund_observation_2001,kapitza_reflection_1933,Reinhardt2020TheoryLight,vanLeeuwen2023FeasibilityBeams,ebel_inelastic_2023,Kempers2024}, or in the confined optical near-fields of laser-excited nanostructures via the scalar potential.\cite{talebi_near-field-mediated_2019,talebi_strong_2020,Hergert2021StrongNanostructures}.
In free space, stimulated interactions appear mainly as longitudinal energy modulation or transverse momentum modulation. The former arises from coherent photon exchange that generates discrete energy sidebands \cite{ebel_inelastic_2023,kozak_ponderomotive_2018}, while the latter, exemplified by the Kapitza–Dirac effect, involves diffraction of electrons by standing-wave light\cite{freimund_observation_2001,kapitza_reflection_1933}. Extending these concepts to structured light, Laguerre–Gaussian modes enable orbital angular momentum transfer \cite{Lloyd2017ElectronMomentum,Grillo2014HighlyHolograms}, producing vortex beams and mode-selective excitations for advanced spectroscopy and imaging.\cite{Koutensky2025Ultrafast4DScanningSTEM}
In the near-field regime, on the other side, optical fields confined to sub-wavelength volumes by plasmonic antennas \cite{Piazza2015SimultaneousNear-field,Vanacore2020Spatio-temporalInteraction}, dielectric metasurfaces, or photonic-crystal cavities enable strong coupling, multi-photon sideband generation, and precise phase matching. Efficient electron–light coupling requires compensation of the inherent momentum mismatch between the electron and the optical field. This can be achieved either by employing dielectric or plasmonic nanostructures that provide an additional momentum through spatial confinement or by guiding the optical mode in appropriately designed subwavelength geometries. 
\par
Shaped electron wavefunctions allow precise control over quantum electrodynamic interactions, scattering processes, Bremsstrahlung emission\cite{Wong2021ControlWavepackets}, control X-ray generation\cite{Wong2024EnhancingWavepackets}, and provide a means to disentangle quantum interference pathways \cite{Lim2023QuantumWavefunctions}. This approach leads to improvements in imaging resolution\cite{Beche2017EfficientImaging,Bucher2024CoherentlyInterferometer}, selective probing \cite{guzzinati_probing_2017,Streshkova2024ElectronNanoscale}, and low-dose imaging\cite{Vanacore2020Spatio-temporalInteraction} to prospective roles in quantum computing \cite{Reinhardt2021FreeElectronQubits}, and high-capacity data transmission \cite{Ropke2021DataModulation}. \par

\begin{table*}[t]
\centering
\footnotesize 
\setlength{\tabcolsep}{5pt} 
\renewcommand{\arraystretch}{1.15} 

\caption{Energy regimes in electron--light interactions.}
\label{tab:energy_regimes}

\begin{tabular}{|L{2.6cm}|L{2.6cm}|L{3.5cm}|L{6.5cm}|}
\hline
\textbf{Regime} & \textbf{Energy Range} & \textbf{Techniques} & \textbf{Physical Considerations} \\
\hline
Ultraslow & $10$--$500\,\mathrm{eV}$ &
PPM, LEED, LEEM~\cite{Storeck2020CDWULEED,Bauer1994LEEM,Bauer2020LEEM,Salancon2019PointSourcePPM,Woste2023UltrafastMicroscope} &
Dominant recoil; low mean free path; high sensitivity to surface potentials. \\
\hline
Low-Energy & $0.1$--$5\,\mathrm{keV}$ &
ULV-SEM~\cite{talebi_strong_2020,ebel_inelastic_2023,talebi_near-field-mediated_2019} &
Recoil dynamics are dominant; high sensitivity to electromagnetic fields. \\
\hline
Medium-Energy & $5$--$30\,\mathrm{keV}$ &
SEM~\cite{shiloh_quantum-coherent_2022,kozak_inelastic_2017,chirita_mihaila_transverse_2022} &
Valid non-recoil approximation for thin films and nanostructures; used for beam shaping. \\
\hline
High-Energy & $>30\,\mathrm{keV}$ &
TEM, STEM, UEM~\cite{feist_quantum_2015,Dahan2020ResonantWavefunction,barwick_photon-induced_2009} &
Relativistic electrons; negligible recoil; standard PINEM framework. \\
\hline
\end{tabular}
\end{table*}

In this review, we focus on the theoretical and experimental progress on stimulated electron–light interactions for free-electron probes with energies below 30 keV (from $10\,\mathrm{eV}$ to $30\,\mathrm{keV}$). As summarized in Table~\ref{tab:energy_regimes}, the relevant techniques cover distinct electron-energy ranges, from low-energy diffraction and microscopy (LEED/LEEM, $\sim 1$--$150\,\mathrm{eV}$ ~\cite{Storeck2020CDWULEED,Bauer1994LEEM,Bauer2020LEEM}) and lensless point-projection approaches (PPM/UPPM, $\sim 50$--$500\,\mathrm{eV}$ \cite{Salancon2019PointSourcePPM,Woste2023UltrafastMicroscope}), to SEM experiments ($\sim 0.1$--$30\,\mathrm{keV}$ ~\cite{talebi_strong_2020,ebel_inelastic_2023,talebi_near-field-mediated_2019,shiloh_quantum-coherent_2022,kozak_inelastic_2017,chirita_mihaila_transverse_2022}). In all of these methods, the observability of stimulated electron–light coupling is governed by electron coherence, synchronization/phase stability, and spatial mode overlap with optically driven fields. At ultralow energies, stray-field sensitivity, and charging can obscure coherent signatures. Whereas the coherence length of a standard LEED optics is typically only $10$--$20\,\mathrm{nm}$, limited by the energy spread $\Delta E$ and beam divergence.\par

We begin with free-space electron–light interactions (Section \ref{Free-Space electron light interactions}), outlining how optical fields enable elastic and inelastic momentum exchange. This discussion contrasts the classical ponderomotive perspective (Section \ref{chapter_freespace_class}) with the quantum description (Section \ref{chapter_freespace_quant}) of stimulated Compton scattering and the Kapitza–Dirac effect.
We then survey near-field mediated interactions (Section \ref{chapter_Near-field mediated electron light interactions}), where optical confinement in plasmonic or dielectric nanostructures overcomes intrinsic momentum mismatch and enables phase-matched coupling, within ultrafast SEM (Section \ref{chapter_Photon-Induced Near-Field Electron Microscopy (PINEM) with Slow Electrons}) and PPM (Section \ref{chapter_Ultrafast Point-Projection Microscopy (UPPM)}) .
Next, we address recoil effects (Section \ref{chapter_recoil}), which become significant for low-energy electrons and strongly influence both energy and momentum transfer. Building on these foundations, we review concepts and methods for longitudinal and transverse electron-wavepacket shaping (Section \ref{beam shaping}), enabling temporally engineered and spatially structured electrons through controlled angular and linear momentum exchange. 
Finally, we discuss emerging pathways such as hybrid optical–electrostatic shaping, laser-based aberration correction, and optical interferometry, which point toward compact, reconfigurable electron–optical platforms. 

\section{Free-space electron light interactions}\label{Free-Space electron light interactions}
\begin{figure}[t]
    \centering
    \includegraphics[width=0.85\linewidth]{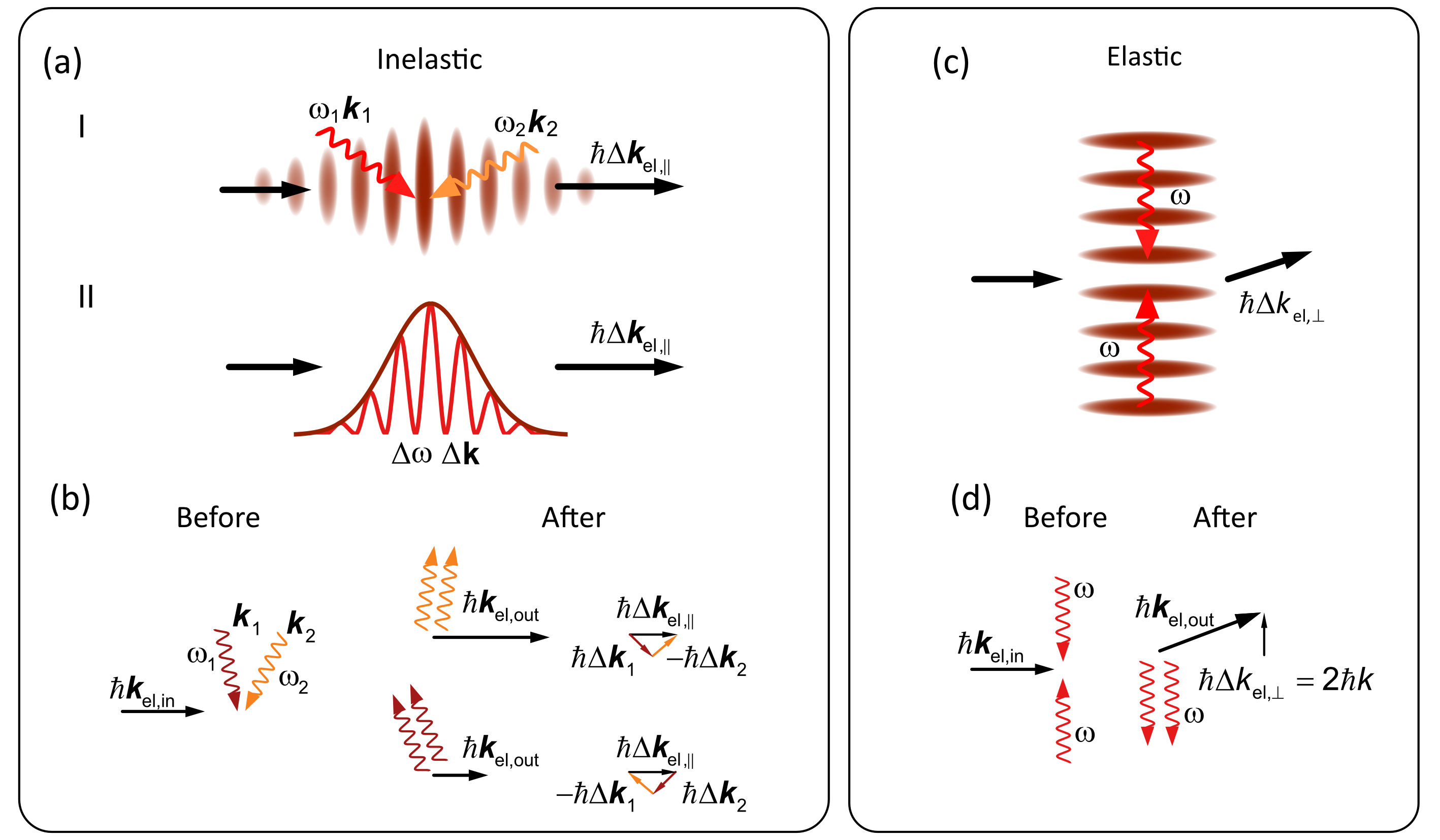}\label{Fig_scattering}
    \caption{Illustration of inelastic (a,b) and elastic (c,d) free electron light interactions. (a,c) depict the ponderomotive potentials enabling inelastic and elastic electron scattering.  Panel (a) I illustrates the ponderomotive potential formed by traveling wave. Panel (a) II illustrates the ponderomotive potential formed by a single laser pulse. (b)  Stimulated Compton scattering (energy gain and loss). (d) Illustration of elastic stimulated Compton scattering as described by the Kapitza-Dirac effect.  }\label{scattering_ill}
\end{figure}

Free-space interactions between electrons and light are in general kinematically forbidden by the simultaneous conservation of energy and momentum. However, when an electron interacts with structured or sufficiently intense electromagnetic fields that provide additional momentum, such as those formed by standing light waves or tightly focused laser pulses, both inelastic and elastic scattering processes become possible (see Figs.~\ref{scattering_ill}a and c). In this regime, the interaction proceeds via higher-order multiphoton processes, the simplest example being stimulated Compton scattering. In this process, two light beams with wave vectors $\mathbf{k}_1$ and $\mathbf{k}_2$ intersect along the trajectory of an electron with initial momentum $\mathbf{p}$. The lowest-energy interaction leads to the electron absorbing a photon from one optical beam, and stimulating emitting a photon via interactions with the second beam, where the momentum conservation of such a process is stated as

\begin{equation}
    \mathbf{p} +\hbar \mathbf{k}_1 = \mathbf{p}' + \hbar \mathbf{k}_2,
\end{equation}
with $\mathbf{p}'$ denoting the electron momentum after the interaction.\\
Thus, the momentum and energy transfers are
\begin{equation}
\label{momentum_conv_1}
    \Delta \mathbf{p} = \mathbf{p}' - \mathbf{p} = \hbar (\mathbf{k}_1 - \mathbf{k}_2) ,
\end{equation}
and 
\begin{equation}
\label{energy_conv_1}
    \Delta E = E(\mathbf{p}') - E(\mathbf{p}) = \hbar(\omega_1 - \omega_2),
\end{equation}
For a small transferred momentum, $\Delta \mathbf{p} \ll \mathbf{p}$, corresponding to the non-recoil approximation, the final energy of the electron can be expanded around its initial momentum $\mathbf{p}$ as
\begin{equation}
\label{taylor_energy}
    E(\mathbf{p}') \approx E(\mathbf{p}) + \frac{\partial E(\mathbf{p})}{\partial \mathbf{p}} (\mathbf{p}'- {\mathbf{p}}).
\end{equation}
After identifying $\partial E/\partial \mathbf{p} = \partial \omega/\partial \mathbf{k} = \mathbf{v}$ as the initial group velocity of the electron, Eq.~\ref{taylor_energy} can be inserted into Eq.~\ref{energy_conv_1}. Using Eq.~\ref{momentum_conv_1} to substitute for the momentum change then yields the following:
\begin{equation}\label{PhaseMatching}
    \omega_1 - \omega_2 = \mathbf{v}\cdot(\mathbf{k}_1 - \mathbf{k}_2),
\end{equation}
which constitutes the phase-matching condition required for inelastic scattering, leading for example, to free-space acceleration of electron beams (Fig.~\ref{scattering_ill}b). 

This expression covers two limiting scenarios. In the general case, it describes the scattering between different beams (Fig. \ref{scattering_ill}d), while in the symmetric case ($\omega_1 = \omega_2,\, \mathbf{k}_1=-\mathbf{k}_2$), it reduces to an elastic geometry, which is known as the Kapitza–Dirac effect\cite{Kapitza1933TheWaves}.\\
These processes can be interpreted from either a classical or a quantum-mechanical perspective. In the classical picture, cycle-averaged ponderomotive forces generated by the intense light field deflect and modulate the electron trajectory. In the quantum picture, the interaction is described as the coherent absorption of a photon with energy $\hbar\omega_1$ and the stimulated emission of a photon with energy $\hbar\omega_2$.

\subsection{Classical description}\label{chapter_freespace_class}
\begin{figure}[t]
    \centering
    \includegraphics[width=1\linewidth]{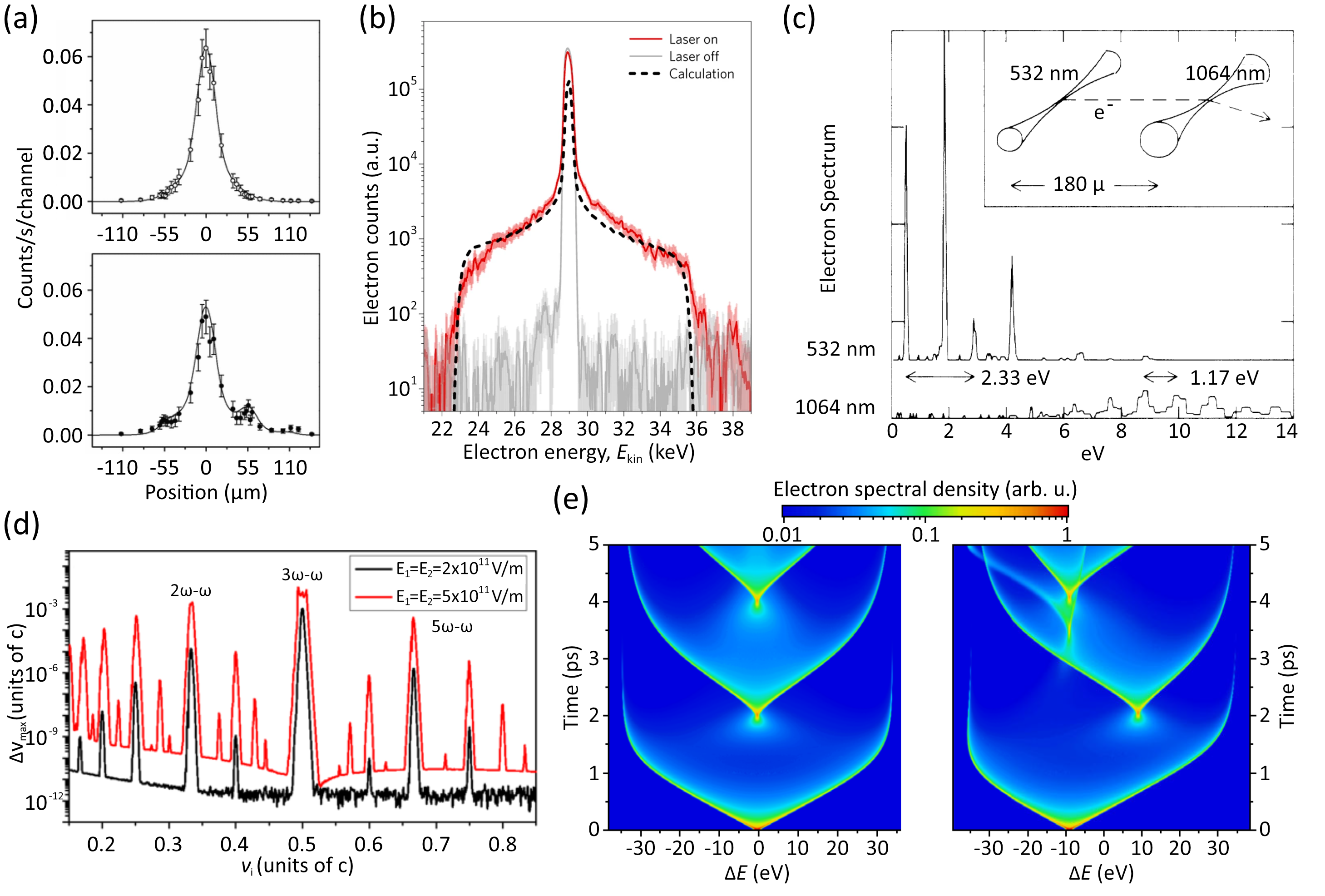}
    \caption{Ponderomotive elastic and inelastic electron light interaction. (a) Experimental observation of elastic electron diffraction into a single diffraction order, as originally proposed By Kapitza and Dirac. Reproduced with Permission ~\cite{Batelaan_2002_Bragg}. Copyright 2002, Physical Review Letters (b) Experimental realization of an inelastic electron light scattering by a traveling ponderomotive grating formed by two inclined laser pulses. Reproduced with Permission ~\cite{kozak_ponderomotive_2018}. Copyright 2018, Nature Physics (c) Inelastic electron scattering from a single laser pulse. Reproduced with Permission ~\cite{bucksbaum_scattering_1987}. Copyright 1987, Physical Review Letters. (d) Higher order nonlinear inelastic electron light scattering resulting into enhanced electron velocity modulation. Reproduced with Permission ~\cite{kozak_nonlinear_2018}. Copyright 2018, Physical Review A. (e) Comparison between synchronous (velocity matched) and asynchronous electron light scattering where the latter results in the formation of narrow linewidth accelerated electrons. Reproduced with Permission ~\cite{kozak_asynchronous_2022} Copyright 2022, Physical Review Letters.}\label{FigPond}
    
\end{figure}
In the classical description, the interaction of free electrons with light fields is controlled by the cycle-averaged force known as the ponderomotive force. This force arises from the quiver motion of an electron in an oscillating electromagnetic field and can be expressed in terms of an effective potential. For a slowly varying monochromatic field with angular frequency $\omega$, this ponderomotive potential is given by~\cite{Bucksbaum1987}:  
\begin{equation}
    U_p=\frac{e^2E^2}{4m\omega^2} ,
\end{equation}
where $E$ denotes the electric-field amplitude, and $m$ and $e$ are the electron mass and charge, respectively.
In a standing optical wave, this ponderomotive potential acquires a periodic spatial profile and therefore acts as a diffraction grating for the electron. Kapitza and Dirac originally predicted that in such a geometry, electrons would undergo elastic diffraction from the optical grating.~\cite{Kapitza1933TheWaves} While their prediction was formulated within a fully quantum–mechanical description of a single–electron wave, the underlying physics can also be captured classically by considering the trajectories of many electrons moving in the cycle–averaged ponderomotive potential.~\cite{Chan1979} In this classical picture, the spatial modulation of the ponderomotive potential leads to angularly separated electron trajectories and ultimately to the formation of a diffraction peak. This Bragg-type diffraction has been experimentally demonstrated by Freimund and Batelaan,~\cite{Batelaan_2002_Bragg} as shown in Fig.~\ref{FigPond}a, and is known as the Bragg-diffraction regime of the Kapitza–Dirac effect.

The inelastic scattering of electrons from light can be achieved by considering the temporal evolution of the ponderomotive potential. An experiment by Bucksbaum et al.~\cite{bucksbaum_scattering_1987} demonstrated that electrons interacting with an intense, pulsed laser field can gain or lose continuous amounts of energy (Fig.~\ref{FigPond}c). Electrons encountering the leading edge of the laser pulse are accelerated, whereas those crossing the trailing edge are decelerated, effectively \textit{surfing} the changing potential of the pulse envelope. However, at the peak of the pulse, the electrons experience an elastic deflection as a result of the symmetric ponderomotive potential.

Building on this single-pulse inelastic electron–light interaction, several follow-up studies demonstrated that this mechanism can be used to temporally compress electron pulses. In these schemes, the ponderomotive interaction imprints a negative velocity chirp on the electron bunch, accelerating the trailing electrons while decelerating the leading ones. Hence, after free propagation, this velocity gradient results in a temporal compression of the electron pulse in the femtosecond and even attosecond regime.~\cite{PhysRevLett.86.5274,doi:10.1073/pnas.0709019104}
This concept has further been extended by employing Gaussian, Hermite–Gaussian, and Laguerre–Gaussian optical modes, which offer enhanced compression capabilities while reducing temporal distortions and improving phase stability.~\cite{wong_all-optical_2015,hilbert_temporal_2009}\par
More recently, the concept of ponderomotive scattering has been extended to traveling optical gratings.~\cite{Kozak2018InelasticInvacuum} In this approach, Kozák et al. generated a moving intensity grating by overlapping two femtosecond laser pulses of different frequencies at non-collinear angles. They demonstrated that the resulting traveling optical potential can be velocity-matched to the electron beam, thereby enabling efficient longitudinal energy transfer (Fig.~\ref{FigPond}b). 
After free propagation, this sinusoidally imprinted velocity modulation evolves into a train of attosecond electron pulses,~\cite{Kozak_Atto_2019} opening pathways toward laser-driven acceleration and attosecond-resolved electron diffraction. 
Kozák et al. furthermore showed that introducing a third laser pulse into the interaction geometry enables simultaneous elastic and inelastic compression, allowing the generation of an isolated attosecond electron pulse at electron energies around 200~keV.~\cite{Kozak_Atto_2019}
\par
A further regime of ponderomotive scattering involves higher-order nonlinear inelastic interactions between free electrons and an optical standing wave, which imprint a high-frequency and purely longitudinal energy–momentum modulation onto the electrons. This nonlinear mechanism, first described by Kozák~\cite{Kozak2018NonlinearWave}, considers a process in which an electron absorbs $N$ photons from one optical wave while emitting $L$ photons into a counter-propagating wave. He showed that by selecting resonant initial electron velocities, $v_i=c(N-L)/(N+L)$, the ponderomotive interaction can be steered into different nonlinear regimes, such as the three-photon (2$\omega$–$\omega$), four-photon (3$\omega$–$\omega$), and six-photon (5$\omega$–$\omega$) processes. These regimes follow the characteristic intensity-scaling laws of $E_1^2E_2$, $E_1^3E_2$, and $E_1^5E_2$, respectively. Such higher-order interactions lead to a strongly enhanced longitudinal velocity modulation of the electron beam (Fig.~\ref{FigPond}d).

The inelastic electron interaction with a traveling ponderomotive grating was further studied by considering an asynchronous interaction between the electron and the optical grating. Where, asynchronous refers to the situation in which the electron velocity cannot fulfill the phase-matching condition required for the stimulated Compton scattering. In the ponderomotive picture, the electron then experiences the traveling potential in an asymmetric manner, which results in a unidirectional energy shift and the formation of electron energy states with a markedly narrow spectral width (Fig.~\ref{FigPond}e).~\cite{Kozak2022AsynchronousWaves}

\subsection{Quantum description}\label{chapter_freespace_quant}
When moving from the classical picture of a point charge interacting with an optical field to a quantum-mechanical description, the interaction must again be framed in terms of stimulated Compton scattering (see Fig.~\ref{scattering_ill}). Although the classical formulation captures the cycle-averaged ponderomotive forces acting on an electron, the Compton picture remains restricted to considerations of energy and momentum conservation alone. A more complete understanding therefore requires treating the electron as a matter wave. In this framework, elastic and inelastic electron–light interactions emerge from the evolution of the electron wavefunction under the time-dependent Schrödinger equation with the minimal-coupling Hamiltonian:

\begin{equation}
i \hbar \frac{\partial}{\partial t} \psi(\vec{r}, t)=\left[-\frac{\hbar^2}{2 m_0} \nabla^2+\frac{e^2}{2 m_0}|\vec{A}(\vec{r}, t)|^2-\frac{i \hbar e}{m_0} \vec{A}(\vec{r}, t) \cdot \vec{\nabla}\right] \psi(\vec{r}, t)\text{ .}
\end{equation}
Within the Coulomb gauge ($\vec{\nabla}\cdot\vec{A}=0$), the scalar potential constitutes the longitudinal field that could be neglected for the transverse free-space wave when optical-fields are tightly focused. $\vec{A}(\vec{r},t)$ is the vector potential, $m_0$ is the electron mass, and $e$ is the electron charge. There are both analytical and numerical approaches to solving this equation within the context of electron-light interactions.~\cite{wolkow_uber_1935,batelaan_colloquium_2007,talebi_interference_2019,konecna_electron_2020}\par
\begin{figure}[t]
    \centering
    \includegraphics[width=1.0\linewidth]{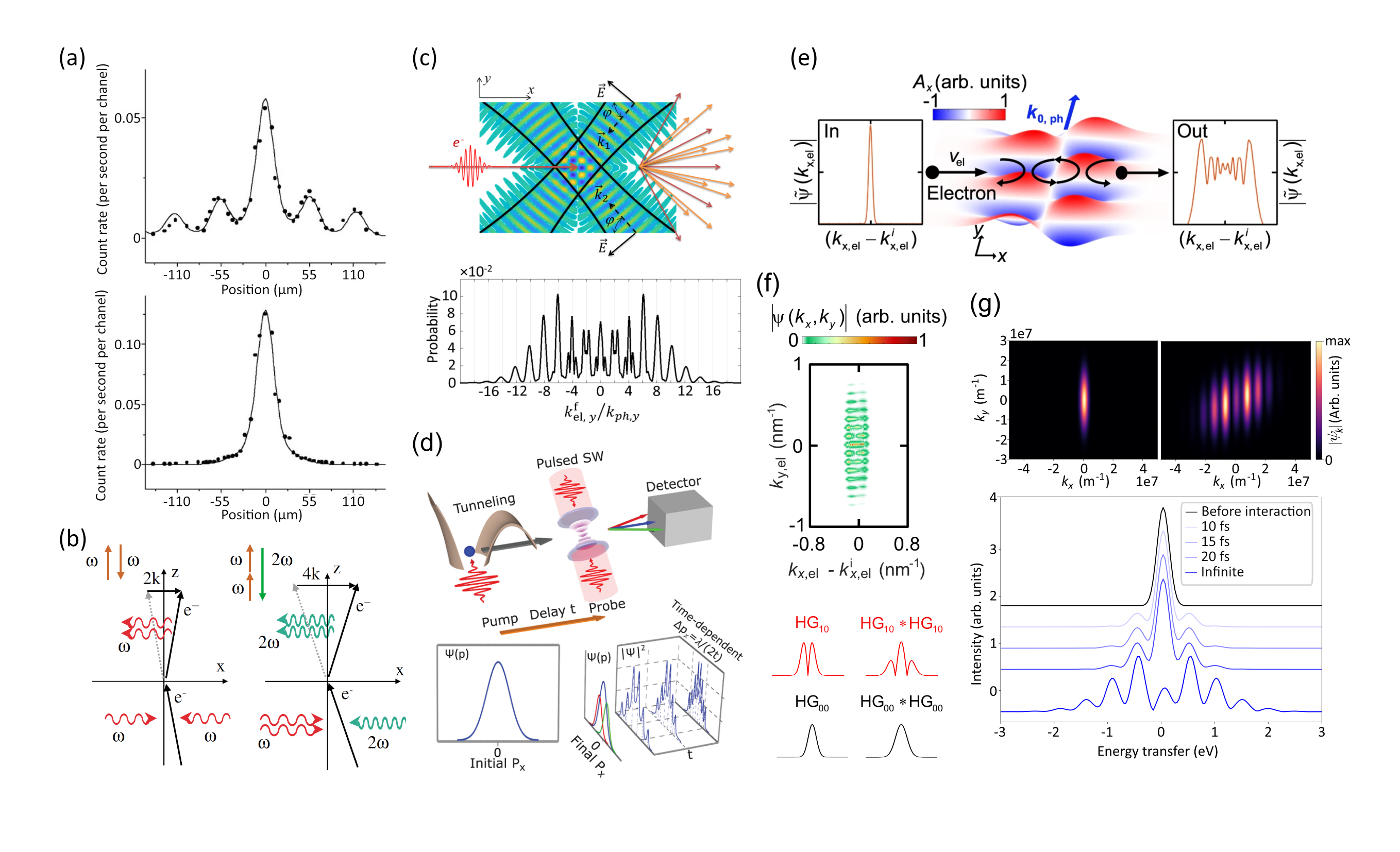}
    \caption{Quantum effects in free space electron light interactions. (a) experimental observation of the Kapitza-Dirac effect, resulting in the diffraction of an electron matter wave from a standing light wave into multiple diffraction orders. Reproduced with Permission ~\cite{freimund_observation_2001}. Copyright 2001, Nature (b) Higher order nonlinear Kapitza-Dirac effect allowing to control the momentum order separation according to the number of involved photons. Reproduced with Permission ~\cite{smirnova_kapitza-dirac_2004}. Copyright 2004, Physical Review Letters (c) Quantum path interferences arising in the Kapitza-Dirac effect when considering an electron wavepacket interacting with the light grating formed by two inclined laser beams. Reproduced with Permission ~\cite{talebi_interference_2019}. Copyright 2019, New Journals of Physics, licensed under CC BY 4.0. (d) Ultrafast Kapitza-Dirac effect with wide momentum bandwidth electron wavepacket leading to time-dependent Kapitza-Dirac diffraction orders. Reproduced with Permission ~\cite{UltrafastKapitza}. Copyright 2024, Science. (e) Inelastic scattering of an electron wavepacket from single pulsed structured light beam, resulting in the formation of energy sidebands. Reproduced with Permission ~\cite{ebel_inelastic_2023}. Copyright 2023, communications physics, licensed under CC BY 4.0 (f) Elastic and inelastic electron scattering from two counter-propagating structured light pulses forming a convoluted momentum distribution enabling control about the energy sideband separation in the final electron energy gain spectra. Reproduced with Permission ~\cite{Ebel2025StructuredWaves}. Copyright 2025, New Journals of Physics, licensed under CC BY 4.0 (g) Momentum space shaping of a Gaussian wave packet by stimulated Compton scattering (upper). Energy transfer diagram for increasing interaction strength (lower).}\label{FigQpond}
\end{figure}

When revisiting the interaction of an electron with the optical potential generated by a standing light wave, the resulting field acts as a periodic grating for the electron matter wave. In this regime, the electron wavefunction is coherently diffracted into discrete momentum orders separated by $2k_\mathrm{ph}$ as illustrated in Fig.~\ref{FigQpond}a.~\cite{batelaan_colloquium_2007}
This phenomenon corresponds to the quantum regime of the Kapitza–Dirac effect, where diffraction arises from the coherent exchange of photon momentum between the electron and the standing optical field. The first experimental observation of this quantum diffraction was reported in 2001 by Freimund et al.~\cite{freimund_observation_2001} who demonstrated multi-order Kapitza–Dirac scattering using a 380~eV electron beam.

The monochromatic standing-wave picture of the Kapitza–Dirac effect was further generalized by Smirnova et al.~\cite{smirnova_kapitza-dirac_2004}, who demonstrated that electron diffraction can also arise from two counter-propagating optical fields of different frequencies. In this configuration, no stationary standing wave is formed; instead, the electron interacts with a field in which multiphoton exchange becomes the dominant mechanism. An electron may absorb $N$ photons from the field of frequency $\omega_{1}$ and stimulate the emission of $M$ photons into the field of frequency $\omega_{2}$, while satisfying the energy-conservation condition $N\omega_{1} = M\omega_{2}$. 
A representative example is shown in Fig.~\ref{FigQpond}b for the case $\omega_{2} = 2\omega_{1}$. Where, the electron absorbs two photons at $\omega_{1}$ and emits a single photon at $\omega_{2}$, preserving energy while receiving a net momentum transfer of $4\hbar k_{\mathrm{ph}}$. 

Motivated by extending electron–light interactions beyond the monochromatic standing-wave geometry, Talebi et al.~\cite{talebi_interference_2019} introduced an elastic scattering configuration in which two inclined laser beams of identical frequency $\omega$ interact with an electron wavepacket. As shown in Fig.~\ref{FigQpond}c, this arrangement produces a standing-wave modulation perpendicular to the electron trajectory, while simultaneously generating a traveling optical pattern along the propagation direction. The resulting optical potential shapes the electron dynamics through an interplay of quantum pathways. In this regime, both the $A^2$ contribution and the $A \cdot \mathbf{k}$ interaction channel of the minimal-coupling Hamiltonian become relevant, allowing population of intermediate momentum states, as illustrated in the bottom panel of Fig.~\ref{FigQpond}c.

The ultrafast Kapitza–Dirac effect was experimentally demonstrated by Lin et al.~\cite{UltrafastKapitza}, in an interaction regime where a free electron, generated by strong-field ionization and characterized by a momentum spread larger than the reciprocal periodicity of the optical grating, interacts with a transient light grating formed by two femtosecond laser pulses. As shown in Fig.~\ref{FigQpond}d, the electron wavepacket acquires time-dependent diffraction features that arise from the self-interference between the unperturbed component of the wavepacket and the momentum-shifted states generated by the Kapitza–Dirac interaction.\par
Similarly to the quantum description of elastic stimulated electron–light interactions, the inelastic process also leads to the formation of discrete energy sidebands in the electron spectrum after the interaction. This effect was recently demonstrated by Tsarev et al.~\cite{tsarev_nonlinear-optical_2023}, who observed sideband formation for 95~keV electrons interacting with two inclined laser beams arranged to satisfy the stimulated Compton scattering condition. Their measurements provide direct experimental confirmation of earlier theoretical predictions~\cite{haroutunian_analogue_1975,Kozak2022AsynchronousWaves}.
Ebel and Talebi~\cite{ebel_inelastic_2023} further demonstrated that inelastic scattering from a single Hermite-Gaussian pulsed laser beam can also generate discrete electron-energy sidebands, as illustrated in Fig.~\ref{FigQpond}e. 
Capturing this process requires going beyond the dipole approximation and incorporating the full spatial dependence of the optical field. 
They demonstrated that the interplay between the temporal pulse envelope and the transverse beam structure creates a ponderomotive potential landscape in which the electron wavepacket undergoes inelastic scattering and self-interference, giving rise to well-resolved energy sidebands. This behavior can be naturally interpreted within the framework of stimulated Compton scattering. The finite spatial extent of the focused pulse introduces longitudinal momentum components, which together with the broadband spectrum of the short driving field, can satisfy the phase-matching conditions required for discrete sideband formation.\par
Within the structured-light framework, two counter-propagating optical pulses with tailored spatial profiles can be configured to induce both elastic Kapitza–Dirac diffraction and inelastic stimulated Compton scattering simultaneously. In this configuration, each diffraction order acquires a characteristic comb-like energy spectrum, as shown in Fig.~\ref{FigQpond}f (upper panel).~\cite{Ebel2025StructuredWaves} By further varying the combination of spatial modes involved in forming the optical field, the longitudinal momentum transfer experienced by the electron wavepacket can be engineered in a controlled manner. Consequently, both the spectral positions and the relative populations of the sidebands become tunable parameters, as illustrated in Fig.~\ref{FigQpond}f (lower panel). \par

The results discussed above highlight that inelastic electron–light interactions mediated by structured optical fields rely critically on phase-matching and are therefore constrained, at optical wavelengths, to relatively slow electrons in the few-keV range. A reduced electron velocity influences physics in two ways: first it relaxes the phase-matching condition and facilitates momentum exchange with optical fields, it simultaneously enhances the interaction time and thus strengthens the overall coupling strength. As the interaction strength scales with the effective temporal overlap between the electron wavepacket and the optical field, lower velocities generally enhance the net modulation.
Fig.~\ref{FigQpond}g illustrates how possible variations in interaction strength, for example, by modifying the interaction time directly shape the electron’s stimulated-Compton energy-gain spectrum. In a recent study, Velasco et al.~\cite{Velasco2025} demonstrated that this limitation can be overcome by employing two counter-propagating continuous-wave Gaussian beams that propagate collinearly with the electron trajectory. This configuration enables millimeter-scale interaction lengths, thereby substantially enhancing the coupling strength and revealing new regimes of free-space electron–light interaction.\par
Beyond the pure question on how to discretely control free electron momentum and energy states in free space by light, there have been efforts to introduce the Kapitza-Dirac blockade, the suppressed sequential climbing of harmonic oscillator states in electron light interaction, to achieve the generation of free electron cat, and non-Gaussian states.~\cite{huang_kapitza-dirac_2021} Furthermore, recent advances have extended the spin-dependent Kapitza–Dirac effect and the electron spin–dependent elastic diffraction~\cite{ahrens_spin_2012} to optical frequencies, potentially opening a pathway to free-electron spin control at low electron energies.~\cite{Tian2024}\par
The classical and quantum descriptions of free-space electron–light interactions have thus far been introduced as distinct regimes. In practice, however, these frameworks are deeply interconnected and represent different physical limits of the same underlying interaction. The transition to the quantum regime is formalized by treating the electron as a wavepacket. The longitudinal extent of the wavepacket, $\Delta z$, is intrinsically linked to the energy spread $\Delta E$ and the delocalization of the electron wavepacket in a coherent manner as in contrast with classical treatment, where an electron ensemble is energetically distributed within a certain statistical model. In an experimental context, this energy spread determines the duration over which the electron maintains a stable phase relationship, reflecting the degree of longitudinal coherence within the pulse. The distinction between classical and quantum pictures can be characterized by comparing the coherence length $\Delta z$ with the spatial periodicity $\lambda$ of the optical field ~\cite{Pan2023WeakInteractions}. When $\Delta z \ll \lambda$, phase coherence across multiple periods of the optical cycle is not maintained; consequently, the interaction is well-described by classical, cycle-averaged ponderomotive forces acting on a point-like particle. Conversely, when $\Delta z \gg \lambda$, interference between different momentum components becomes observable, necessitating a quantum-mechanical description in terms of stimulated Compton scattering and matter-wave diffraction. Recent numerical studies by Kuchař et al. have rigorously compared these classical and quantum frameworks for moving ponderomotive potentials, demonstrating their consistency within their respective regimes of validity ~\cite{Kuchar2025Ponderomotive}. As an experimentally observable clear distinction as for which model is suited to which experimental condition, the energy quantization of the electron wavepacket within the stimulated Compton scattering interaction regimes, as well as the transverse quantization within the classical Kapitza-Dirac effect are distinguished. Both of these phenomena demand quantum mechanical modeling, as the classical treatment does not model the coherent bunching of the electron wavepacket as a result of quantum interactions between light and delocalized electron wavepackts.\par
In summary, here we discussed how free-space electron–light interactions, which are typically forbidden in a vacuum due to energy–momentum conservation criteria, can become possible when structured or sufficiently intense optical fields provide the requisite momentum mismatch, through nonlinear interaction regimes. This condition can be described from two complementary perspectives: a classical picture and a quantum framework. The transition between these two descriptions is fundamentally governed by the ratio of the electron wavepacket extent ($\Delta z$) to the optical wavelength ($\lambda$).\par
Although free-space interactions reveal the basic structure of stimulated electron–photon coupling, phase mismatch and limited interaction lengths restrict the achievable modulation with slow electrons. These limitations motivate the use of nanostructure mediated near-fields, which provide the additional momentum required for efficient coupling and are therefore central to the slow-electron regime discussed in the next section.

\section{Near-field mediated electron light interactions}\label{chapter_Near-field mediated electron light interactions}
Among the various methods for strengthening electron–light interactions, near-field coupling provides an especially effective approach by exploiting optical modes confined to subwavelength volumes~\cite{Vanacore2020Spatio-temporalInteraction}. \par
These near-fields possess wavevectors much larger than those of free-space propagating lights, allowing efficient compensation of the intrinsic energy–momentum mismatch. When an electron beam passes near a laser-excited nanostructure, it interacts with the evanescent or localized optical modes of the material~\cite{talebi_strong_2020,Kozak2018UltrafastNanostructures}, enabling efficient exchange of energy and momentum with the optical field. Such interactions are experimentally realized in electron microscopes and point-projection microscopes, where ultrafast excitation schemes provide direct access to the underlying dynamics.  
These properties make slow-electron near-field interactions not only a powerful method for high-resolution imaging but also a phase-resolved spectroscopy tool for mapping complex photonic fields at the nanoscale\cite{Koutensky2025Ultrafast4DScanningSTEM}. However, for electrons with kinetic energies of only a few electronvolts, nanostructures must have dimensions of only a few nanometers to achieve sufficient confinement for phase matching and efficient electron–light coupling. Indeed, efficient energy exchange requires a synchronization (phase-matching) condition between the electron wavepacket and the optical field, ensuring that the electron wavepacket samples a well-defined optical phase during the interaction. For the interaction to remain coherent, the effective interaction time must be synchronized with the oscillation period of the field\cite{talebi_strong_2020}. At even lower electron velocities, the need for greater optical momentum compensation becomes important, demanding proportionally smaller feature sizes. However, a small electron velocity widens the electron beam waist and requires a more localized near-field to fulfill the phase-match condition, thereby
reducing the chance of electron-photon interaction. This scaling imposes practical limitations on coupling efficiency and motivates the design of advanced nanophotonic structures capable of maintaining phase matching under extreme slow-electron conditions.

\subsection{Photon-induced near-field electron microscopy with slow electrons}\label{chapter_Photon-Induced Near-Field Electron Microscopy (PINEM) with Slow Electrons}

\begin{figure}[t]
    \centering
    \includegraphics[width=1\linewidth]{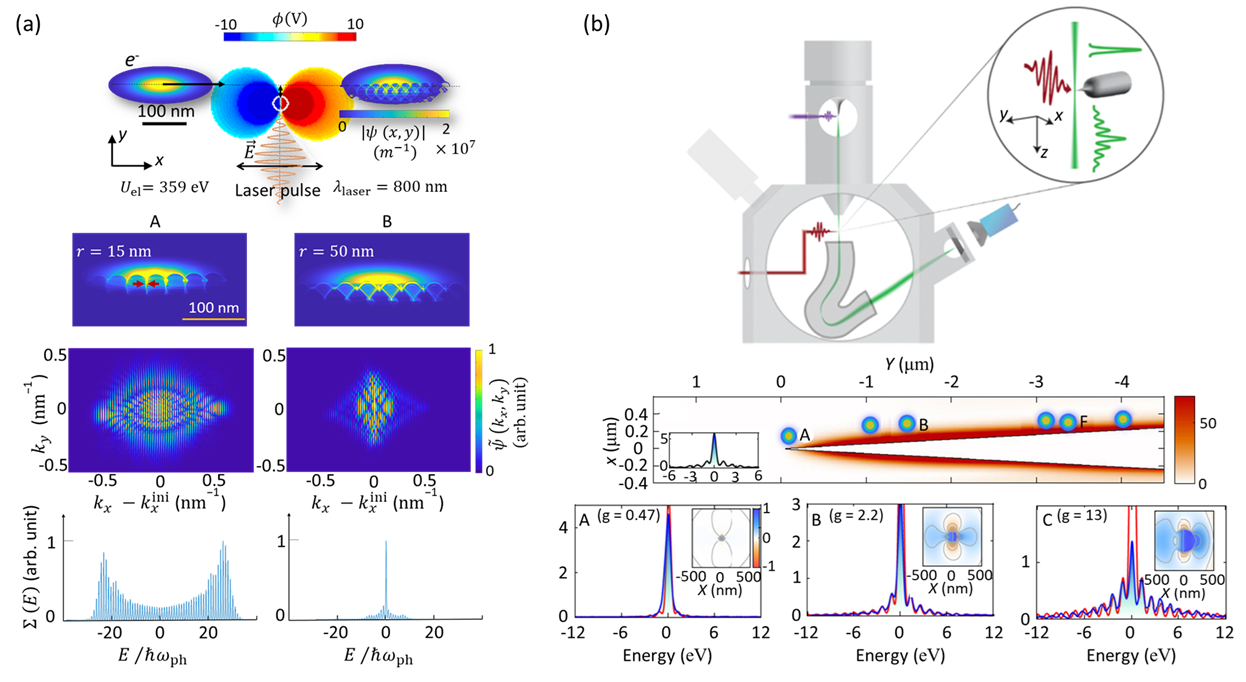}
    \caption{ Theoretical and experimental demonstrations of PINEM inside SEM. (a) First-principles simulations showing electron modulation by sub-keV electrons, controlled by the strength of its coupling with optical near-fields, which results on amplitude and phase modulation of a Gaussian electron wavepacket. Hence it experiences attosecond bunching in real space, and longitudinal energy exchange, and elastic transverse diffraction in momentum representation, after the interactions. Reproduced with Permission \cite{talebi_strong_2020}. Copyright 2020, Physical Review Letters. (b) Quantum-coherent coupling of electrons to laser-driven near-fields inside a SEM, with a custom-designed magnetic energy analyzer, and the number and amplitude of the photon orders in PINEM spectra vary as a function of electron beam position. Where \(g\) indicates the simulated near-field coupling parameter, and panels (A-C) present PINEM spectra recorded at different positions along the tungsten needle tip (colored circles), demonstrating position-dependent coupling strength at an electron beam energy of 17.4 keV. The inset illustrates the the cross section of the real part of the simulated normalized electric field component. Reproduced with Permission \cite{shiloh_quantum-coherent_2022}. Copyright 2022, Physical Review Letters}
    \label{fig 5}
\end{figure}
An electron beam traversing the optical field of a laser-illuminated nanostructure can undergo inelastic scattering and exchange an integer number of \(\ell\) quanta of photons with frequency \(\omega_{\mathrm{ph}}\). Therefore, the resulting electron spectrum features a discrete comb like ladder of sidebands spaced by \(\hbar\omega_{\mathrm{ph}}\), with the probability of occupying the \(\ell\)-th sideband given by:
\begin{equation}
P_\ell = |J_\ell(|g|)|^2 \, ,
\end{equation}
where \(J_\ell\) is the Bessel function of the first kind and \(g\) is the dimensionless coupling parameter:
\begin{equation}
g = \left( \frac{e}{\hbar \omega_{\mathrm{ph}}} \right) 
\int_{-\infty}^{\infty} dx' \, \tilde{E}_x(x',y) \,
e^{-i x' \omega_{\mathrm{ph}} / v_{\mathrm{el}}} ,
\end{equation}
\(\tilde{E}_x\) is the longitudinal near-field component along the electron trajectory, with \(v_{\mathrm{el}}\) and \(\omega_{\mathrm{ph}}\) representing the electron velocity and the angular frequency of the photon, respectively. \cite{park_photon-induced_2010,barwick_photon-induced_2009}. In its simplest form, this discrete energy bunching is described within a one-dimensional model that neglects the electron recoil \cite{park_photon-induced_2010}. \par
Within the realm of PINEM, non-recoil approximation serves as an analytical tool to determine the coupling strength. Indeed, this $g$-parameter, even in the realm of slow-electron and light interactions, has remained as the main parameter quantifying the strength of the interaction with light. In other words, whereas the $g$-parameter is universally identified as the “coupling strength” it only quantifies the strength within the non-recoil approximation and is valid for fast electrons. For the case of ultraslow electron wavepackets, neither the  $g$-parameter nor the non-recoil approximation remains valid. Here, full-wave numerical simulations, which do not remain within the constraints of non-recoil approximation, have to be used to model the interactions and observations such as near-field mediated Kapitza-Dirac effect.\par
 However, in the weak-coupling regime between the field and the electron \((eg/\hbar \omega_{\mathrm{ph}} \ll 1)\), the spectrum is dominated by first-order sidebands and can be described perturbatively \cite{park_photon-induced_2010}; higher-order absorption and emission peaks appear with a rapidly decreasing intensity relative to the zero-loss peak \cite{park_photon-induced_2010} (Fig. \ref{fig 5}a, lower plot in panel B). Whereas, in the strong-coupling regime \((|g| \gg 1)\), coherent energy exchange with the optical field significantly modifies the electron wavepacket, depleting the zero-loss peak and resulting in populating multiple higher-order sidebands (Fig. \ref{fig 5}a, lower plot in panel A) \cite{feist_quantum_2015}. This regime not only enhances coupling efficiency but also enables the generation of non-classical states, including electron-photon entanglement \cite{kfir_entanglements_2019}.\par
Over the past two decades, the quantum nature of free-electron wavepackets has been experimentally demonstrated through the transformation of TEMs into ultrafast, quantum-coherent instruments. Until recently, most studies were confined to TEMs, where typically only one or two photon–electron interaction sites could be achieved. \par
Talebi~\cite{talebi_strong_2020} extended this frontier by investigating the interaction between low-energy electrons and laser-excited photons of the gold nanorod using time-dependent Maxwell-Schrödinger simulations from first-principles. Her work highlighted the decisive influence of synchronicity condition between near-field oscillation, and electron wavepacket propagation $\lambda_{\mathrm{ph}} \, v_{\mathrm{el}} = 2r$ and revealed that, in addition to inelastic photon absorption and emission (PINEM sidebands), slow electrons interacting with plasmonic near-fields can undergo elastic photoinduced diffraction (Fig.~\ref{fig 5}a). Her simulations further show that the strength of the interactions governs both the amplitude and phase modulation of a Gaussian electron wavepacket.\cite{talebi_strong_2020} Where the amplitude modulation leads to attosecond temporal bunching in real space, and the phase modulation manifests itself as a population of discrete momentum orders along the longitudinal and transverse directions in momentum space. Under phase-matched conditions (Fig. \ref{fig 5}a, A), the interaction produces electron bunching at the attosecond-scale, sharply defined diffraction patterns, and strong-coupling sideband structures, while the lack of synchronization in larger multimode nanorods (Fig. \ref{fig 5}a, B) leads to a reduced coupling efficiency and diffuse spectral features.\par
Further numerical studies have shown that by tailoring the geometry, polarization, and topology of the nanostructure~\cite{chahshouri_tailoring_2023}, as well as by adjusting the phase offset between sequential near-field components and employing phase-locked gating, the momentum transfer to the electron can be deterministically controlled~\cite{chahshouri_tailoring_2023,chahshouri_numerical_2023}. This control enables Ramsey-like interference schemes, where sequential, phase-correlated interactions act as temporal beam splitters and recombiners for the electron wavepacket.
The conventional theoretical framework describes materials through their dielectric functions; however, this classical electrodynamic approach often overestimates the coupling strength in ultrasmall metallic structures, where quantum tunneling alters field confinement.\par
\begin{figure}[t]
    \centering
    \includegraphics[width=1.0\linewidth]{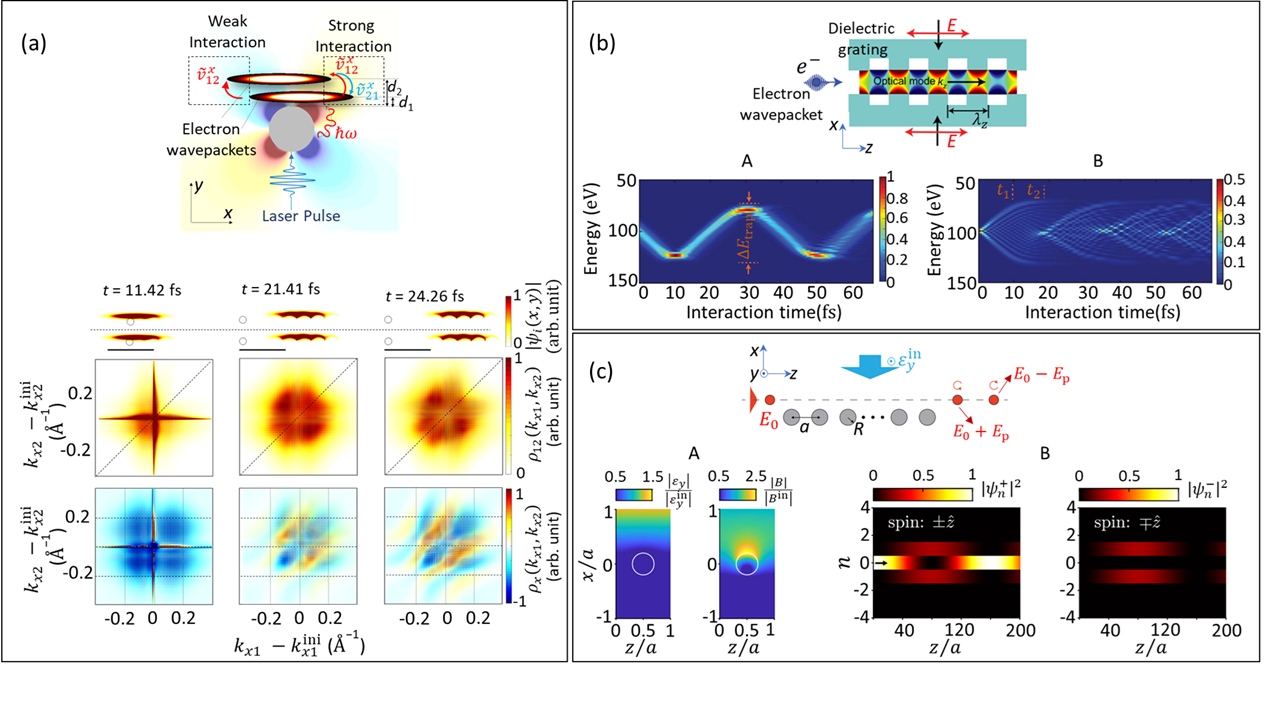}
    \caption{Emerging frontiers of stimulated interactions of slow electrons with light. (a) Exchange-mediated correlations in a spin-polarized two-electron system interacting with the plasmonic near-fields of a gold nanorod excited by a laser pulse. The first and second wavepackets traverse the near-field zone at distances of 5 nm and 10 nm from the nanorod surface. Middle and lower maps show the total and exchange components of density matrices at each time, respectively. Reproduced with Permission \cite{Talebi2021Exchange-mediatedInteractions}. Copyright 2021, New Journals of Physics, licensed under CC BY 4.0. (b) Phase-matched interactions of 20–200 eV electrons with optical fields. Schematic of a possible experimental setup in which an electron beam interacts with a phase-matched optical mode of a dielectric grating illuminated from both sides by a laser. Lower maps, simulated momentum sideband populations of the electron as a function of interaction time for two initial wavepacket conditions: a Gaussian wavepacket (A) and a plane wave (B), both centered at an initial kinetic energy of 100 eV, and photon energy is \(\hbar\omega_{\mathrm{ph}}\)= 1.54 eV. Reproduced with Permission \cite{eldar_self-trapping_2022}. Copyright 2024, Physical Review Letters. (c) Spin polarization of free electrons in optical near-fields. 
A transverse-electric (TE) laser beam excites a nanowire array, generating near-fields that contain both electric and magnetic components with a relative phase delay. 
A free electron propagating parallel to the array interacts with these fields, resulting in spin-dependent transitions between energy sidebands. 
(A) Simulated distributions of the electric (right) and magnetic (left) near-fields surrounding a nanowire array excited by a plane wave with a photon energy of \(\hbar\omega_{\mathrm{ph}}\) = 1.24 eV. 
(B) Spin-resolved probability densities for spin-preserving (\(|\psi_{n}^{+}|^{2}\)) and spin-flip (\(|\psi_{n}^{-}|^{2}\)) transitions, plotted as colored ribbons across discrete energy levels, illustrate how optical excitation of a nanostructure can induce spin polarization in free electrons through coupled electric–magnetic near-field interactions. Reproduced with Permission ~\cite{Pan2023PolarizingFieldsb}. Copyright 2023, Physical Review Letters}
    \label{Fig 6}
\end{figure}
In continuous SEM-type operation (e.g., Schottky sources), the measured energy spread is highly sensitive to operational conditions, as reducing the beam current or emitter temperature can significantly narrow the distribution from several eV to sub-eV levels \cite{Kim1997EnergyDistributionsZrOW}. Typical energy spreads vary by emitter type: thermionic LaB$_6$ ($\sim$1--2~eV), thermionic W (1.5--3~eV), Schottky FEG (0.5--1~eV), and cold field-emission (0.2--0.4~eV).\cite{Egerton2009ElectronTEM,Arbouet2018UTEM_HistoricalDevelopment} However, in pulsed operation, increasing the electron density per pulse introduces space-charge effects. These include transverse broadening and longitudinal energy broadening known as the Boersch effect \cite{Kuwahara2016BoerschEffectPicosecond} where stochastic Coulomb repulsion degrades both temporal and spatial coherence. Since ultrafast needle/nanotip sources generate electrons in a highly confined space-time volume, even a small number of electrons can interact strongly through stochastic Coulomb repulsion.\cite{Meier2023FewElectronCorrelations} \par
So far, most studies in this realm have been conducted numerically, as the lack of a suitable energy analyzer for SEMs has limited experimental functionality. Recently, however, Shiloh et al. demonstrated a magnetic prism–based energy analyzer capable of operating at electron energies above 10 keV\cite{shiloh_quantum-coherent_2022}, achieving an energy resolution better than 500 meV. Their experiments \cite{shiloh_quantum-coherent_2022} demonstrated that low-energy electrons can act as highly sensitive probes of optical near-field excitations. They observed a quantum-coherent coupling between electrons and photon in a SEM at sub relativistic energies down to 10.4 keV. Due to the spacious and easily reconfigurable geometry of the SEM chambers, such setups can accommodate extended or cascaded optical arrangements. Their experiments revealed a multipole composition that depended on
the distance from the tungsten tip apex by moving the electron impact position along the shaft, which defines the
strength of the PINEM effect, providing the first direct evidence of quantum-coherent coupling of slow electrons to laser-driven near-fields inside an SEM (Fig. \ref{fig 5}b). These effects result in observable asymmetries and modifications in the PINEM sidebands, providing a clear theoretical signature of quantum-coherent electron–light coupling at sub-keV energies. \par 
Recent theoretical work has further shown that two electrons interacting with the same optical near-field can result in exchange-mediated correlations and dephasing which lead to entanglement-like features and collective quantum phenomena~\cite{Talebi2021Exchange-mediatedInteractions}. As illustrated in Fig.~\ref{Fig 6}a, two spin-polarized electron wavepackets propagating near a gold nanorod at distances of 5 nm and 10 nm from its surface, where both interact with the laser-induced plasmonic near-field. In the weak-interaction regime, the phase of the electron closer to the nanostructure is strongly modified and partially transferred to the second electron through exchange, while in the strong-interaction regime, a correlated exchange of phase information produces pronounced modulation in the energy–momentum distributions of both electrons. The density-matrix maps in Fig.~\ref{Fig 6}a highlight how these exchange contributions evolve over time, transferring the phase of the first electron onto the second. Consequently, the quantum state of one electron directly influences the scattering probabilities of the other via their mutual coupling to the optical field. \par 
Surface plasmon polariton (SPP) can also be used not only for spectral modulation but also for direct spatial control of free-electron beams. By tailoring field symmetry, electrons can be compressed, deflected, or coherently shaped on the nanometer and femtosecond scales \cite{Fan2025SpatialInteractions}. Where a symmetric field distribution can produce opposing momentum kicks that focus or squeeze the electron wavepacket, whereas asymmetric excitation leads to a net deflection.\par
Krüger et al. reported a new regime of slow-electron (\(\sim 20{-}200~\text{eV}\)) light interactions, in which electrons interacting with intense optical near-fields no longer exhibit the conventional PINEM ladder of equidistant sidebands. Instead, resonant coupling in this regime leads to a strong confinement of the electron energy spectrum arising from the nonvanishing curvature of the electron dispersion. This curvature acts as an effective quadratic trapping potential in the energy domain, limiting the quantum random walk of the electrons and enabling self-trapping into a small set of discrete energy states (Fig.~\ref{Fig 6}b)~\cite{eldar_self-trapping_2022}. Their simulations further reveal a confined, oscillatory spectral evolution where, for a localized Gaussian wavepacket, the dynamics follow a Lorentzian-like trajectory reminiscent of a charged particle in an oscillating field, and for a plane-wave electron, the spectrum first broadens into a superposition state and then abruptly collapses into a single dominant sideband (Fig.~\ref{Fig 6}b A, and B, respectively).
\par
Pan and Xu~\cite{Pan2023PolarizingFieldsb} recently proposed that slow electrons that interact with transverse-electric (TE) optical near-fields in nanostructures can become spin-polarized by exploiting strong inelastic scattering within phase-matched optical modes. In this process, the two spin components of an initially unpolarized electron beam aligned parallel and antiparallel to the electric field undergo spin-flip transitions and are inelastically scattered into different energy states, providing an energy-domain analog of the Stern–Gerlach experiment. The underlying mechanism, illustrated in Fig.~\ref{Fig 6}c, involves an optical near-field that excites a nanowire array and generates transversely structured near-fields containing electric and magnetic components. A free electron propagating parallel to the array interacts with these fields, resulting in spin-dependent transitions between energy sidebands. The calculated spin-resolved probability densities, $|\psi_{n}^{+}|^{2}$ and $|\psi_{n}^{-}|^{2}$, demonstrate how optical excitation of a nanostructure can induce spin polarization in free electrons via coupled electric–magnetic near-field interactions.\par
In a summary, here we discussed near-field–mediated electron–light interactions as a robust method for achieving strong and controllable modulation of slow-electron wavepackets. In this regime, localized optical near-fields confined to subwavelength volumes provide large wavevectors, thereby enabling efficient energy–momentum exchange when an electron traverses the vicinity of a laser-driven nanostructure. This coupling is reflected in the PINEM spectrum, where the electron acquires discrete energy gain and loss sidebands. The interaction strength is quantified by the coupling parameter g, which distinguishes between two regimes: weak coupling, dominated by low-order sidebands, and strong coupling, where numerous photon orders emerge, the zero-loss peak is depleted, and the wavepacket is reshaped through combined amplitude and phase modulation.\par
This section highlighted furthermore, emerging directions in near-field electron–light coupling that extend beyond the standard PINEM framework. First, when two electrons interact with the same plasmonic near-field, exchange-mediated coupling can transfer phase information between wavepackets, generating correlated, entanglement-like features in their energy–momentum response. Second, in the slow-electron regime (20–200 eV), in which resonant, phase-matched coupling can confine the spectral dynamics and induce self-trapping in the energy domain rather than a standard equidistant sideband ladder; and third, spin-dependent near-field interactions, where coupled electromagnetic fields can drive spin-flip transitions, offering a potential route toward the control of free-electron spin.

\subsection{Ultrafast point-projection microscopy}\label{chapter_Ultrafast Point-Projection Microscopy (UPPM)}
\begin{figure}[t]
    \centering
    \includegraphics[width=0.9\linewidth]{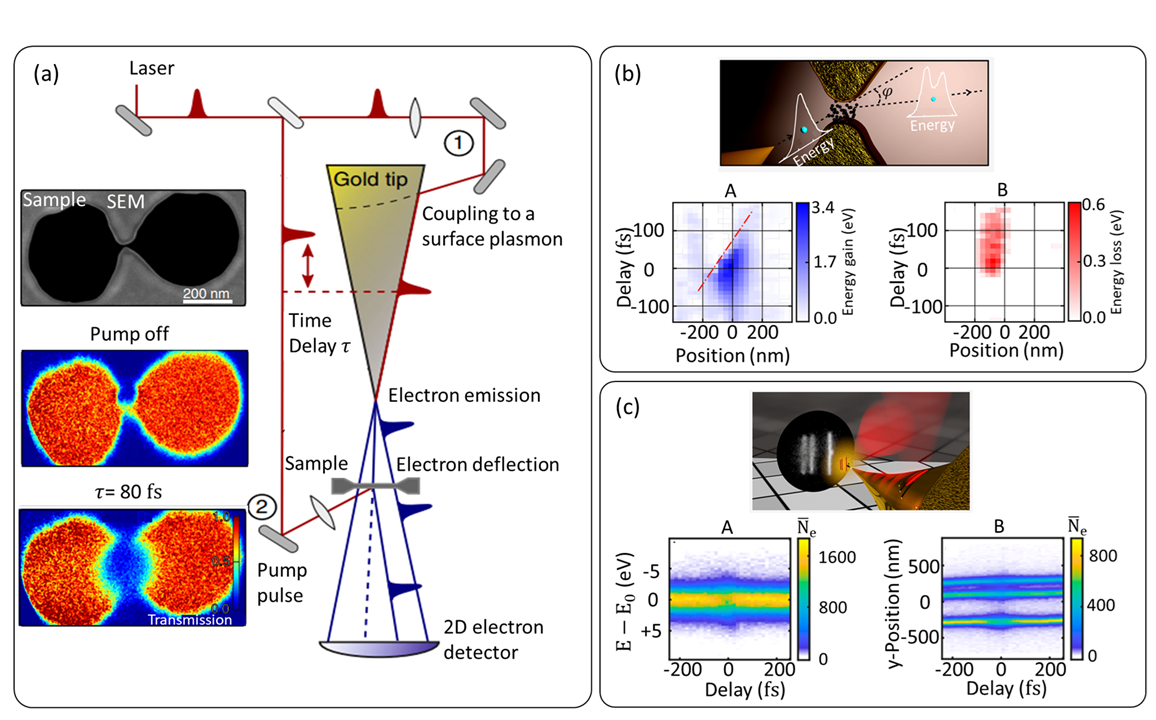}
    \caption{Interaction of ultrafast point-projection electrons with optically excited nanostructures. Femtosecond laser pulses excite plasmonic nano antennas, while ultrafast low-energy electrons probe the dynamics of charge and plasmonic excitation. (a) Laser-driven photoemission and charge-separation dynamics in plasmonic nanoantennas. Reproduced with Permission \cite{Vogelsang2018ObservingMicroscopy}. Copyright 2018, Light: Science \&\ Applications. (b) Transient electric fields created by photoemission from a nanogap antenna, causing probe-electron deflection and acceleration/deceleration. Reproduced with Permission \cite{Hergert2021ProbingMicroscopy}. Copyright 2021, Acs Photonics. (c) Ultrafast coupling of optical near-fields to low-Energy electrons. (A) Kinetic-energy distribution of probe electrons, and (B) deflection, spatially averaged over the area of a Yagi–Uda antenna, plotted as a function of the time delay between the optical pump and electron probe. The electron–near-field interaction is confined to approximately 50 fs around time zero, with spectra centered at 80 eV. Reproduced with Permission ~\cite{Woste2023UltrafastMicroscope}. Copyright 2023, Nano Letters, licensed under CC BY 4.0}
    \label{fig 4}
\end{figure}
With UPPM~\cite{Vogelsang2018ObservingMicroscopy,Thomas2013ProbingScale,Hoff2017TracingLaserpulses} one can achieve a slow energy regime (50-500 eV range) suitable for measuring some of the theoretical studies mentioned in this review \cite{Hergert2021ProbingMicroscopy,Vogelsang2018Plasmonic-Nanofocusing-BasedHolography} and enables time-resolved diffraction, imaging\cite{Paarmann2012CoherentStudy}, and interferometry. In this setup, a sharp metallic nanotip is illuminated by a femtosecond laser pulse to trigger ultrafast photoemission. The resulting electron pulses are intrinsically coherent and highly divergent, producing a lensless projection\cite{Hergert2021ProbingMicroscopy,Woste2023UltrafastMicroscope}, thereby providing a pristine view of electron-light interactions. In this scheme, tip sample separations of tens to hundreds of micrometers ensure ultrashort propagation times and control magnification\cite{Hergert2021StrongNanostructures}.\par
Using this method, Vogelsang et al. probed charge dynamics in nanostructures \cite{Vogelsang2018ObservingMicroscopy}. They generated femtosecond, low-energy electron pulses by plasmon nanofocusing on a sharp gold taper and used them to probe ultrafast photoemissions from a
nanometer-sized plasmonic antenna (Fig. \ref{fig 4}a). They showed that when the antenna was optically excited, the photoemitted electron cloud in the gap region deflected the probe electrons, due to Coulomb interactions. With a combined spatial and temporal resolution of 20 nm and 25 fs, this experiment visualized the spatio-temporal expansion of the photoemitted electron cloud and revealed the accompanying charge separation in the metal\cite{Vogelsang2018ObservingMicroscopy}.\par
Building on this foundation, Hergert et al.\cite{Hergert2021ProbingMicroscopy} extended this approach by equipping UPEM with an energy- and position-resolved delay-line detector. This detector enabled visualization of kinetic-energy changes and transverse deflections of probe electrons caused by longitudinal and transverse components of the local optical field. They further showed that because low-energy electrons experience longer interaction times with localized fields, they experience larger angular deflections compared to fast electrons. Using 80–100 eV probe electrons that interact with the transient fields of a laser-excited nanogap antenna, they observed clear signatures of acceleration (up to +5 eV) in negative delays, deceleration (–1 eV) in positive delays, and characteristic spectral splitting around time zero (Fig.~\ref{fig 4}b). For large negative delays, probe electrons pass the nanoresonator before the photoemitted electron cloud forms and thus remain unaffected, whereas at smaller negative delays, Coulomb attraction to the emerging cloud accelerates them toward the detector. At zero delay, Coulomb repulsion between cloud and probe electrons simultaneously accelerates faster probe electrons and decelerates slower ones, leading to broadening and eventual splitting of the kinetic-energy spectrum.\par 
Most recently, Wöste et al.\cite{Woste2023UltrafastMicroscope} demonstrated the first direct  coupling of low-energy electrons to optical near-fields, rather than merely to photoemitted charge clouds where phase-matching between electrons and near-fields is achieved due to spatial confinement of the antenna near-field. In their experiment, 100 eV electrons with sub-50 fs duration interacted with the localized near-fields of a nanometer-sized Yagi–Uda antenna excited by 20 fs near-infrared laser pulses (Fig.~\ref{fig 4}c). Phase modulation of the electrons by transverse-field components resulted in transient deflection, whereas longitudinal near-field components caused broadening of the kinetic energy distribution. Unlike in conventional PINEM experiments with swift electrons, discrete photon sidebands were not observed in this work because of the finite energy spread of the slow electrons (about 3 eV), which is larger than the photon energy of the near-infrared field, therefore, deteriorating the visibility of coherent PINEM sidebands. Instead, coherent spectral broadening and delay-dependent energy shifts revealed the underlying interaction dynamics, as shown by acceleration/deceleration and deflection dynamics in Fig.~\ref{fig 4}c.\par
Beyond femtosecond resolution, advances in strong-field nanophysics have shown that photoemission from sharp metallic tips can be controlled on the attosecond timescale using carrier envelope phase-stable laser pulses~\cite{Kruger2011AttosecondTip}. Recent studies have demonstrated ultra-nonlinear, few-electron photoemission bursts with sub-femtosecond confinement~\cite{Hergert2024Ultra-NonlinearNanotapers} and suppression of rescattering by moderate bias fields, yielding subcycle electron pulses suitable for ultrafast PPM~\cite{Hergert2025QuenchingFields}. Incorporating such attosecond control into PPM could enhance its temporal resolution, enabling direct access to the dynamics of subcycle electrons and paving the way for attosecond point-projection electron microscopy.\par
This section described Ultrafast Point-Projection Microscopy (UPPM) as a lensless methodology operating in the slow-electron regime (50–500 eV), enabling time-resolved imaging, diffraction, and the investigation of laser-induced near-field dynamics. Within this framework, laser-excited tips and nanoantennas generate transient near-fields and photoemitted charge clouds; these interact with probe electrons to induce delay-dependent acceleration, deceleration, and deflection. This approach highlights recent progress toward direct coherent coupling of low-energy electrons to optical near-fields, extending beyond purely classical Coulomb-driven effects.

\begin{figure}[htbp]
    \centering
    \includegraphics[width=0.8\linewidth]{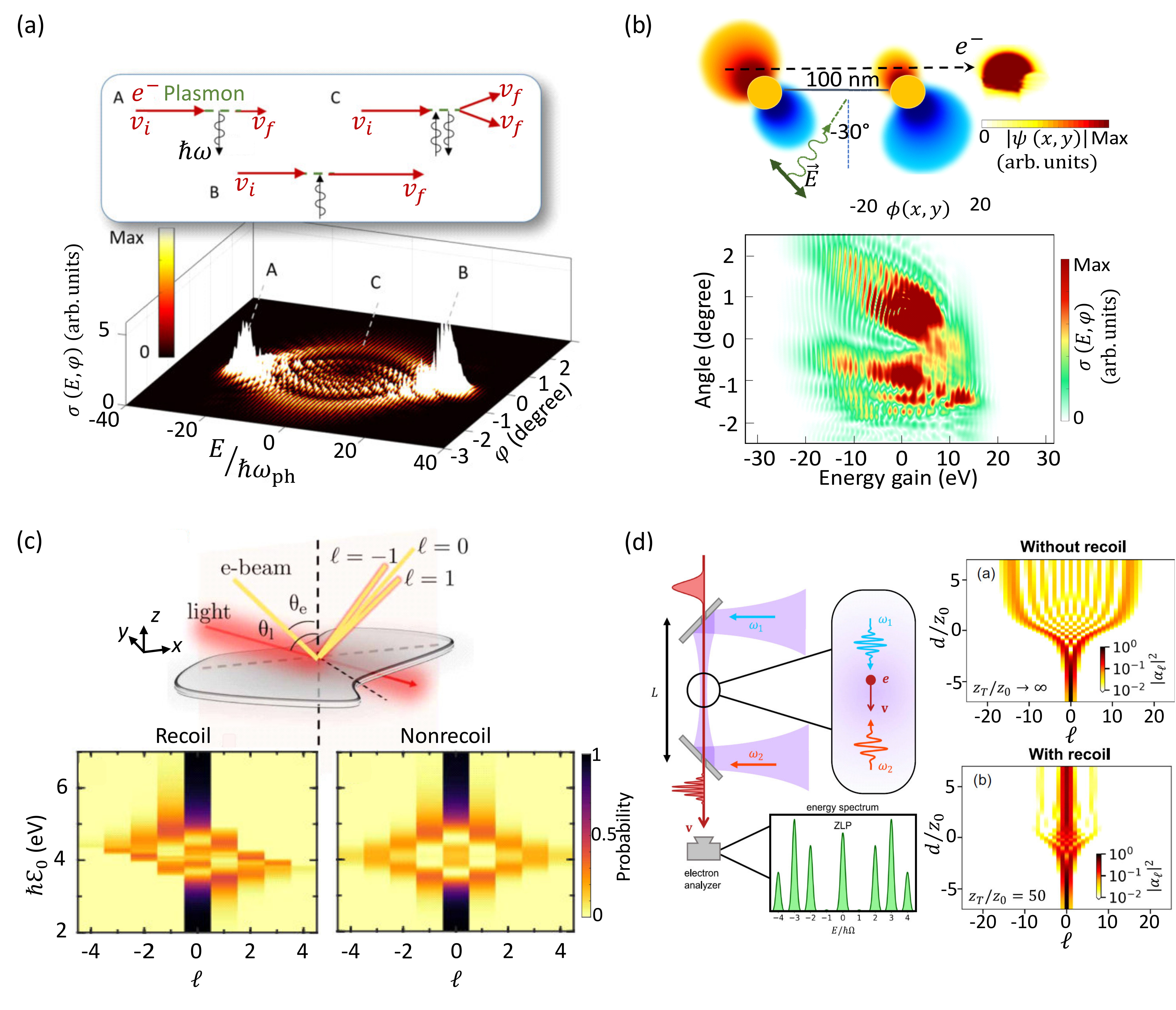}
    \caption{Recoil effects in the interaction of slow electrons with optical near-fields. (a) First-principles simulations by Talebi reveal that for sub-keV electrons interacting with plasmonic nanostructures, recoil becomes a decisive factor, producing asymmetric PINEM sidebands, and elastic diffraction orders. The electrons, with an initial kinetic energy of 400 eV, have transverse and longitudinal FWHM broadening of 2 nm and 20 nm, respectively. The laser pulse features a central wavelength of 800 nm, and a temporal broadening of 16 fs. Reproduced with Permission \cite{talebi_strong_2020}. Copyright 2020, Physical Review Letters. (b) Numerical studies with slow electrons interacting with dual near-field with obliquely illuminated laser. (Top) Amplitude modulation of the electron beam, (bottom) Inelastic scattering cross section map demonstrated recoil-induced sideband asymmetry and transverse deflections.The laser pulse has a central wavelength of 700 nm, and a temporal FWHM broadening of 18 fs. The electron beam is initialized with a central kinetic energy of 600 eV, with longitudinal and transverse FWHM broadening of 56 nm and 36 nm, respectively. Reproduced with Permission \cite{chahshouri_numerical_2023}. Copyright 2023, Scientific Reports, licensed under CC BY 4.0.(c) Schematic of a slow electron beam interacting with a polariton-supporting surface under oblique laser illumination (top). Simulated energy–momentum spectra for an electron undergoing total reflection at a polariton-supporting interface, comparing the recoil regime (left), which produces asymmetric sidebands due to finite momentum exchange, with the non-recoil limit (right), where the distribution remains symmetric. Reproduced with Permission ~\cite{Synanidis2024QuantumLight}. Copyright 2024, Science Advances.(d) (left) Proposed interaction geometry allowing for inelastic free-space electron-light interactions mediated through stimulated Compton scattering with a long interaction region. (right) Evolution of the electron energy gain spectra without (Top) and with (Bottom) consideration of recoil. Reproduced with Permission ~\cite{Velasco2025}. Copyright 2025, Physical Review Letters.}
    \label{Fig 7}
\end{figure}

\section{Recoil effects}\label{chapter_recoil}

In many early descriptions of stimulated inelastic electron--light interactions, including much of the photon-induced near-field electron microscopy studies, the non-recoil approximation was assumed. This approximation, which states that the longitudinal and transverse momentum exchanged between the photon and the electron are small~\cite{Talebi2018Electron-lightInterferometry,RoquesCarmes2023,Huang2023QuantumLattices}, is well suited for relativistic electron beams. In this limit, the electron trajectory is treated as perfectly straight and unaffected by the interaction, and the coupling parameter predicts symmetric energy sidebands corresponding to photon absorption and emission.  
\par
Although classical assumption successfully describes electron energy-loss spectroscopy and PINEM spectra for highly accelerated electrons, it may break down in certain regimes of electron--photon coupling. At low electron energies, high photon energies (e.g., X-rays or $\gamma$ rays), or near velocity thresholds~\cite{RoquesCarmes2023,Talebi2023TunableRecoil,Huang2023QuantumLattices}, recoil effects can induce substantial momentum exchange. Such effects cannot be captured by traditional non-recoil models, motivating the development of theoretical frameworks that explicitly account for recoil in the slow-electron regime. In these conditions, the momentum exchange between the electron and the optical field becomes comparable to the total momentum of the electron, and the longitudinal and transverse recoil components jointly modify both the kinematics and the quantum phases of the interaction. These effects give rise to observable phenomena such as asymmetric sideband intensities, gain-loss imbalance, kinematic thresholds, and phase shifts\cite{Talebi2023TunableRecoil,RoquesCarmes2023,Talebi2018Electron-lightInterferometry}.\par
Talebi provided one of the first comprehensive treatments of this breakdown by solving the full Maxwell–Schrödinger equations for an electron wave packet interacting with plasmonic nanostructures. She introduced the concept of recoil engineering, showing how sideband asymmetry and phase shifts could be tuned through control of electron velocity, near-field confinement, and wave-packet geometry (Fig. \ref{Fig 7}a)\cite{talebi_strong_2020}. Direct comparisons between recoil-inclusive and adiabatic simulations revealed systematic breaking of the gain-loss symmetry, shifts in sideband positions, and velocity-dependent spectral distortions. Beyond longitudinal recoil, transverse recoil\cite{Shi2024TransverseRadiation} where the near-field imparts momentum perpendicular to the propagation direction is also reported  for slow beams, which leads to measurable beam deflection and modified diffraction patterns \cite{talebi_strong_2020}.\par 
By tuning parameters such as electron energy, pulse duration, laser illumination direction, sample geometry, and near-field topology \cite{chahshouri_tailoring_2023}, recoil can be harnessed to engineer asymmetry, enhance phase-matched coupling, and tailor beam profiles. However, the presence of two distinct near-fields further breaks the recoil symmetry, resulting in an asymmetric recoil of the electron wavefunction in the momentum space (Fig. \ref{Fig 7}b). \cite{chahshouri_numerical_2023}
Synanidis and co-workers~\cite{Synanidis2024QuantumLight} investigated how slow electrons, with kinetic energies of only a few electronvolts, scatter from surface polaritons. Their simulations revealed spectra that show abrupt cutoffs and backscattering channels that are classically forbidden. When surface-scattered electrons interacted with an obliquely incident field, the breaking of translational symmetry opened additional coupling pathways absent in the non-recoil model. Figure~\ref{Fig 7}c compares the recoil and non-recoil regimes in low-energy electron–light interactions, where the exchange of quantized momentum sidebands ($\ell = -1, 0, 1$) corresponds to photon absorption or emission. The calculated electron energy–momentum spectra show that in the recoil regime, finite momentum transfer breaks the gain–loss (energy–momentum) symmetry, while in the non-recoil limit the sidebands remain symmetric about the central energy peak.\par
Recoil effects are not limited to PINEM-type experiments. They also play an important role in free-space electron–light interactions within the regime of stimulated Compton scattering. Fig. \ref{Fig 7}d illustrates how electron recoil influences strong free-space interactions in a geometry that enables millimeter-scale interaction lengths.
This geometry causes the electron’s momentum to change during photon exchange, breaking the ideal phase-matching condition of stimulated Compton scattering (see eq. \ref{PhaseMatching}). As a result, the energy spectrum develops a cutoff in the number of observable sidebands, and the spectral shape deviates significantly from the nonrecoil prediction. At the same time, recoil enhances temporal electron compression by inducing velocity-dependent phase shifts between sidebands, leading to stronger pulse focusing and higher temporal coherence.~\cite{Velasco2025}
\par
 Together, these studies establish that recoil effects in stimulated slow-electron interactions are not minor corrections, but defining features of the coupling regime. By treating recoil as a controllable parameter through electron energy, wavepacket duration, interaction geometry, and near-field confinement, it becomes possible to engineer asymmetry, enhance phase-matched energy exchange, and tailor spatial beam profiles. This richer physics offers a pathway to ultrafast, quantum-coherent electron–photon coupling regimes that remain inaccessible to beams constrained by the nonrecoil approximation. However, by correlating 200 keV electrons with coincident transition-radiation photons from a silicon membrane, Preimesberger et al. showed that each emission imparts a measurable momentum kick, with the electron’s distribution reflecting the photon’s angle and wavelength \cite{Preimesberger2025ExploringElectrons}. Although conventional energy-loss or cathodoluminescence measurements mix many processes together \cite{polman_electron-beam_2019}, coincidence filtering isolates the transition radiation channel, revealing that each photon emission event is accompanied by a tiny but measurable recoil of the electron. The electron’s momentum distribution mirrors the photon’s emission angle and wavelength, confirming energy–momentum conservation at the single-event level \cite{GarciadeAbajo2010OpticalMicroscopy}. This approach not only enhances weak radiative signals, but also establishes a pathway to explore quantum correlations and entanglement between free electrons and photons\cite{Henke2025ProbingEraser}.\par
 This section highlighted recoil effects in stimulated electron–light interactions and clarifies the limitations of the widely used non-recoil approximation. While neglecting photon-induced momentum exchange is adequate for high-energy electrons, this approximation breaks down for slow electrons, large photon energies, and within phase-matched interaction scenarios. In this regime, recoil alters both the kinematics and the accumulated phases, producing signatures such as gain–loss imbalance, asymmetric sidebands, and transverse deflection that are not captured by non-recoil models. Consequently, recoil is not merely a correction, but serves as a tunable control parameter through the adjustment of electron energy, wavepacket duration, interaction geometry, and near-field confinement\cite{chahshouri_tailoring_2023}.

 \section{Longitudinal and transverse beam shaping}\label{beam shaping}
 \begin{figure}[H]
    \centering
    \includegraphics[width=1\linewidth]{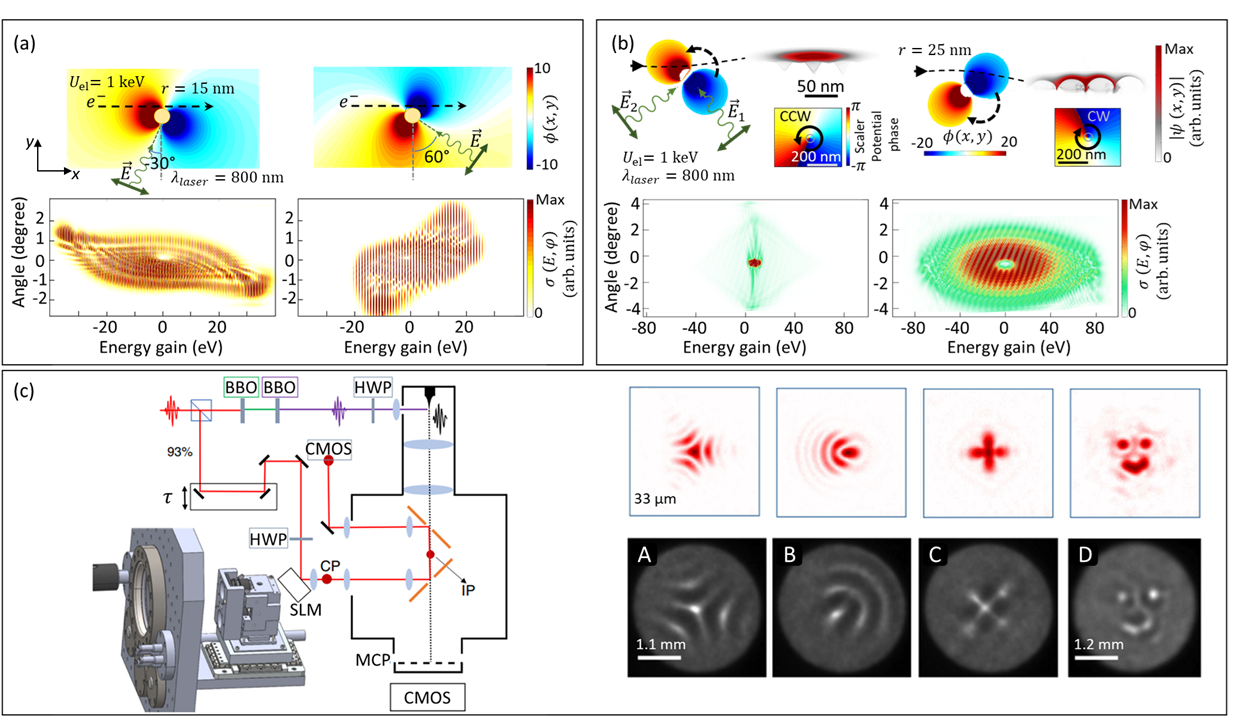}
    \caption{Shaping of slow-electron wavepackets.
(a) Simulated energy–momentum distributions of electrons interacting with dipolar plasmonic near-fields under oblique laser excitation at $-30^{\circ}$ and $60^{\circ}$, respectively. The energy exchange and transverse momentum modulation follow the spatial profile of the near-field from the electron’s perspective, leading to polarization-dependent acceleration and deceleration pathways. The electron beam has a kinetic energy of 1 keV, with longitudinal and transverse FWHM broadening of 40 nm and 5 nm, respectively. The laser pulse is characterized by a central wavelength of $\lambda = 800$ nm, and a 5 fs temporal FWHM broadening. Reproduced with Permission \cite{chahshouri_tailoring_2023}. Copyright 2022, New Journal of Physics, licensed under CC BY 4.0. (b) Electron interactions with circulating plasmonic fields generated by two orthogonally polarized laser pulses with a $\pm\pi/2$ phase delay produce counterclockwise (CCW) and clockwise (CW) rotating near-fields, enabling controlled angular momentum transfer. The handedness of the excitation determines the direction of transverse recoil and the resulting momentum sideband structure.The electron beam has an initial kinetic energy of 1 keV, with a longitudinal and transverse FWHM broadening of 132 nm and 15 nm, respectively. The laser pulses are characterized by a central wavelength of 800 nm, and a temporal FWHM broadening of 21 fs. Reproduced with Permission \cite{Chahshouri2025UltrafastBeams}. Copyright 2025, Nanophotonics, licensed under CC BY 4.0. (c) Experimental setup and demonstration of programmable ponderomotive electron-beam shaping.
(Left) A femtosecond laser (1035 nm, 280 fs, 40 µJ) is frequency-doubled and directed onto a spatial light modulator (SLM) to imprint programmable phase profiles. The shaped laser field is characterized in the conjugate plane (CP) using a CMOS camera and then relayed into the modified SEM chamber, where the electron beam interacts with the counter-propagating light in the interaction plane (IP). The chamber door enables efficient in- and outcoupling of the shaped fields, while an MCP–CMOS detector records the electron signal.
(Right) Optical intensity patterns at the CP (top, scale bar 33 µm) and corresponding electron distributions after interaction (bottom, scale bars 1.1–1.2 mm). The SLM introduces phase patterns corresponding to (A,B) trefoil and coma aberrations, (C) astigmatism, and (D) a “smiley” generated by a Gerchberg–Saxton algorithm. Reprinted from. \cite{chirita_mihaila_transverse_2022}. Copyright 2022, Physical Review X.
}
    \label{Fig 8}
\end{figure}
 
The spatio-temporal phase shaping of the electron wavepacket plays a fundamental role in the advancement of electron microscopy and quantum technologies. It enables improvements in imaging resolution \cite{Vanacore2020Spatio-temporalInteraction}, selective probing of material excitations \cite{Guzzinati2017ProbingBeams}, quantum information processing, and high-efficiency data transmission \cite{Ropke2021DataModulation}, while coherent control of quantum wave functions has led to breakthroughs in bond-selective chemistry \cite{Shen2014TowardsSurfaces}, and ultrafast control of plasmonic excitations \cite{Stockman2008UltrafastControl}.
So far, conventional approaches such as nanofabricated phase masks~\cite{Larocque2018TwistedElectrons,Bliokh2017TheoryStates,Verbeeck2012ASTEM,Roitman2021ShapingFilms}, magnetic-field configurations~\cite{Bliokh2012ElectronStates}, and electrostatic phase plates~\cite{Schattschneider2012NovelBeams} have been employed to manipulate electron wavepackets. However, these methods are inherently limited in terms of speed, dynamic tunability, and the ability to induce simultaneous longitudinal and transverse phase modulation.
 \par
 Within ultrafast scheme, light fields play a role analogous to spatial light modulators (SLMs) in optics, while they provide a dynamic, reconfigurable way to manipulate both the longitudinal and transverse phases of an electron beam. This analogy highlights the versatility of optical control, in contrast to static nanofabricated masks or fixed electrostatic lenses, and sets the stage for treating the electron as a programmable quantum object. \par
Since longitudinal control modifies the electron’s temporal envelope and energy–time correlation, by imprinting an energy chirp using a synchronized optical near-field and then allowing dispersive propagation, electron pulses can be shortened from the picosecond or femtosecond time scale down to the sub-femtosecond or even attosecond domain \cite{Morimoto2018DiffractionTrains}. Indeed, the quantized exchange of energy and momentum between electron and photon enables precise temporal shaping of electron pulses\cite{garcia_de_abajo_optical_2021}. Such short pulses enable coherent phase locking to the optical field and increase the sensitivity to obtain depth information from the interactions. Furthermore, periodic temporal structuring can generate attosecond electron pulse trains with repetition rates set by the optical carrier frequency \cite{Morimoto2020Single-CycleElectrons}, offering a path to stroboscopic probing of ultrafast nanophotonic dynamics with high spectral selectivity\cite{baum_4d_2009,morimoto_diffraction_2017,Morimoto2018_AttosecondTrains}. 
The transverse beam shaping, on the other hand, governs the spatial phase distribution of the electron wavefront, which can be achieved through transverse optical modes of phase-structured near-fields.\par
For slow electrons, the direction of a linearly polarized laser pulse plays a decisive role in spatio-temporal wavepacket shaping. As shown in Fig. \ref{Fig 8}a \cite{chahshouri_tailoring_2023}, the energy exchange and transverse momentum modulation follow the spatial profile of the plasmonic near-field from the electron’s perspective. For instance, when a laser pulse illuminates the nanostructure at a $60^{\circ}$ angle, selective detection of electrons at deflection angles ($\phi > 2^{\circ}$ or $\phi < -2^{\circ}$) through a mechanical slit enables controlled acceleration and deceleration of bunched electron wavepackets. Engineering near-fields through the direction of laser illumination, combined with controlled phase delays in multiple interactions systems, provides a versatile mechanism for manipulating electron wavepackets and selectively populating specific momentum orders\cite{chahshouri_numerical_2023}.\par
Laser-induced circulating near-fields can further control electron modulation by imparting angular momentum in addition to linear momentum to the electrons. Plasmonic rotors, generated either by two orthogonally linearly polarized pulses with $\pm\frac{\pi}{2}$ phase delay or by a circularly polarized laser, produce rotating electromagnetic fields whose circulation direction determines the modulation of the electron wavepacket (Fig.~\ref{Fig 8}b). As demonstrated in Fig.~\ref{Fig 8}b, the handedness of the excitation governs the direction of recoil and the transferred angular momentum, establishing the rotation direction as a fundamental tuning parameter. Moreover, a Doppler-like shift in the frequency of the near-field in the electron’s rest frame controls the coupling strength and, consequently, the spatio-temporal shaping of the electron beam~\cite{Chahshouri2025UltrafastBeams}.\par
\begin{figure}[H]
    \centering
    \includegraphics[width=1\linewidth]{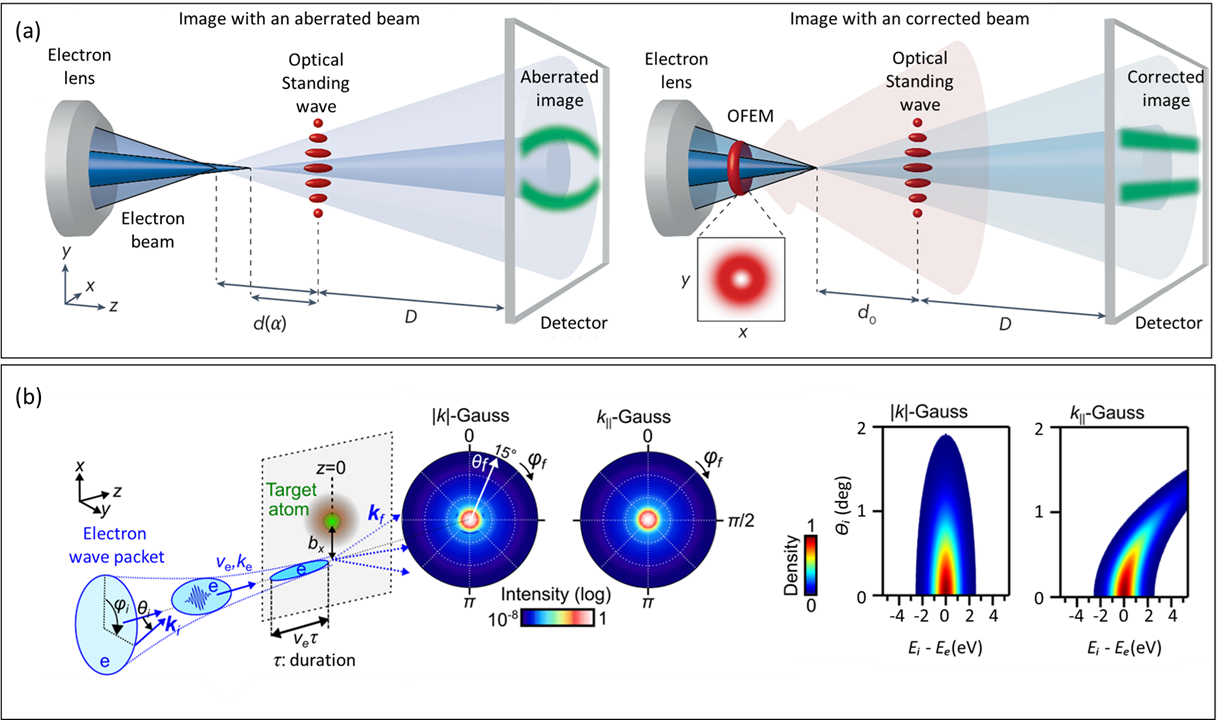}
    \caption{Application of shaped electron beams.
(a) Light-based spherical aberration correction. A Laguerre–Gaussian (LG) laser beam applied upstream of the focus generates an optical field electron mirror (OFEM) that compensates angle-dependent focusing, restoring constant magnification and undistorted fringes. Left panel, image with an aberrated beam, and right panel, aberrated corrected image. Reproduced by permission \cite{ChiritaMihaila2025LightBasedElectronAberrationCorrector}. Copyright 2025, Nature Photonics, licensed under CC BY 4.0. (b) Scattering-asymmetry control with ultrafast shaped electron wave packet. An incident electron wavepacket with initial momentum $\mathbf{k}_i$ and velocity $v_e$ interacts with a target atom or nanostructure, leading to a scattered state . The angular and energy–momentum distributions are compared for two momentum-space wavepacket models: the isotropic $|k|$-Gauss, where the electron has a symmetric momentum spread, and the directional $k_{\parallel}$-Gauss, where the momentum distribution is elongated along the propagation axis. The latter results in asymmetric energy–angle spectra due to the preferential momentum transfer along the electron trajectory. Reproduced by permission\cite{Morimoto2025}. Copyright 2025, New Journal of Physics, licensed under CC BY 4.0.
}
    \label{Fig 9}
\end{figure}
Further experiments employing programmable ponderomotive phase modulation have enabled dynamic shaping of free-electron wavefunctions through interaction with spatially structured femtosecond laser fields. As illustrated in Fig.~\ref{Fig 8}c, a SLM is used to tailor the transverse phase of the laser wavefront, which is then projected into the interaction plane of a SEM chamber~\cite{chirita_mihaila_transverse_2022}. By adjusting the SLM phase map, Mihaila et al.~\cite{chirita_mihaila_transverse_2022} demonstrated electron patterns analogous to optical modes (Fig.~\ref{Fig 8}c, right side, upper) such as trefoil, coma, and astigmatism, as well as user-defined structures like a smiley face, directly mapping the optical field profile onto the electron intensity distribution. They further showed that programming the laser’s transverse intensity profile via SLM can emulate conventional optical elements, including convex and concave lenses, astigmatism compensators, and higher-order phase correctors. The resulting ponderomotive potential imprints the designed phase pattern onto the passing electron beam, enabling the real-time generation of aberration-corrected or custom-shaped intensity distributions.\par
Recently, a light-based electron aberration corrector was used in SEM, in which a Laguerre-Gaussian (LG) laser beam acts as an optical field electron mirror (OFEM) to compensate for spherical aberrations in a focused electron beam (Fig. \ref{Fig 9}a). Mihaila et al.\cite{ChiritaMihaila2025LightBasedElectronAberrationCorrector} demonstrated that this approach corrects the dominant third-order spherical aberration that usually limits electron optics. In this scheme, the LG mode reshapes the electron phase front so that rays with different convergence angles are refocused to a focal point, thereby restoring constant magnification and eliminating image distortions. \par
Morimoto et al. \cite{Morimoto2025} theoretically explore the use of shaped electron wavepackets to actively control scattering asymmetry at the atomic scale. Considering a tightly focused electron beam and a single-atom target displaced off the beam axis, they demonstrate that the angular distribution of elastically scattered electrons becomes asymmetric. By tailoring the wavepacket’s momentum and temporal profile (on the attosecond--picosecond scale), the sign and magnitude of the scattering asymmetry can be tuned, as reflected in the angle-dependent intensity patterns shown in Fig.~\ref{Fig 9}b, which differ for isotropic $\lvert k \rvert$-Gaussian and directional $k_{\parallel}$-Gaussian wavepackets.\par
This section explained how shaped electron beams transform the electron wavepacket into a controllable and programmable quantum probe, encompassing both longitudinal (energy–time) and transverse (spatial) degrees of freedom. Optical and near-field phase modulation offers a high-speed, reconfigurable, and concise alternative to static electron-optical elements. It can impose an energy chirp to enable pulse compression or the formation of pulse trains, while simultaneously sculpting the transverse wavefront by controlling momentum orders or transferring angular momentum. Recent progress in programmable ponderomotive modulation is discussed, where SLM-shaped light is used to imprint user-defined phase profiles for beam formation and aberration correction. The primary applications of shaped electrons are then outlined, including light-based aberration correction, the generation of custom beam modes, and enhanced imaging fidelity.

\section{Conclusion and outlook}\label{Conclusion and outlook}

Over the past decades, light-driven control of electron beams has evolved into a powerful bridge connecting nanophotonics, ultrafast spectroscopy, and quantum optics. Moreover, electron beams which once were regarded as imaging probes can now be coherently shaped, accelerated through optical near-fields, and used as sensitive probes of energy- and charge-transfer dynamics. At the same time, advances in hybrid optical electrostatic designs and in-vacuum photonic interfaces are expected to enhance coupling efficiency and scalability, paving the way for long-term stable and high-precision control of electron–light interactions. \par
In conclusion, the diverse pathways towards stimulated electron–light interactions constitute a set of complementary regimes, distinguished by their momentum-exchange mechanisms, experimental observables, and theoretical requirements. Free-space interactions, being fundamentally constrained by multiple photon and nonlinear interactions, necessitate the use of standing or traveling optical gratings to provide the required momentum mismatch. In this regime, cycle-averaged ponderomotive models provide an intuitive description of classical elastic diffraction and electron-beam acceleration, whereas quantum scattering frameworks become indispensable when electronic coherence is preserved and discrete momentum exchange dominates, as in the case of stimulated Compton scattering and Kapitza-Dirac effect. \par
Near-field coupling represents the most efficient interaction modality enabling direct and linear single-photon and electron interactions, as evanescent modes naturally provide the large wavevectors and strong spatial localization required for the generation of PINEM sidebands. While the weak-coupling regime is ideally suited for field mapping and spectroscopy, the strong-coupling regime enables advanced wavepacket reshaping and time-domain control, albeit with increased sensitivity to geometric constraints, dissipative losses, and phase stability. In the slow-electron regime, UPPM implementations provide a route to explore the regimes of mainly recoil-dominated and extreme quantum pathway interferences, due to the sensitivity of slow electrons to the electromagnetic interactions. \par
Beyond platform-specific distinctions, the validity of the non-recoil approximation separates high-energy TEM experiments from recoil-sensitive slow-electron regimes. Recoil-inclusive treatments are essential when the momentum exchange and phase accumulation become dominant due to long interaction times or phase-matching considerations. Such interactions produce characteristic asymmetric sidebands, phase shifts, and transverse kicks, while simultaneously introducing novel control parameters through electron energy tuning, wavepacket preparation, and interaction geometry.\par
Finally, electron-beam shaping serves as the bridge between these interaction regimes and their practical applications. While static phase elements and conventional electron optics offer robust shaping, optical and near-field methodologies provide reconfigurable, ultrafast control over both the energy–time longitudinal structure and the transverse wavefront. This expanded programmability facilitates transformative applications, including pulse compression, tailored on-demand beam shapes, light-based aberration correction, and engineered scattering outcomes at the atomic scale.\par

Advances in electron beam shaping have transformed the role of light in electron microscopy from a passive illumination source into an active, programmable modulator of electron quantum states. This optical programmability makes it possible to compress electron pulses to attosecond durations, correct optical aberrations in real time, and transfer linear and angular momentum with high precision. These advances bring electron microscopy closer to the regime of fully coherent and adaptive beam control, where the electron wavefunction can be shaped on demand for mode-resolved electron energy-loss spectroscopy and cathodoluminescence \cite{Karnieli2023QuantumElectrons}.\par
Beyond improving imaging and spectroscopic performance, such control opens the door to quantum-state engineering and phase-sensitive interferometry with free electrons. By tailoring both the longitudinal and transverse phases, one can encode and manipulate quantum information directly within the electron’s wavefunction, realizing a new class of experiments in which the coherence and phase of the electron become measurable resources and use this system of solid-state quantum systems for sensing. \par
Despite rapid progress, the degree of control achievable in electron modulation is still bounded by a small set of practical limitations that ultimately determine how far photon–electron interactions can be pushed experimentally. A first constraint is energy–momentum readout for resolving fine sideband structures and diffraction features. In the very low-energy regime (100 eV), time-of-flight detection combined with an MCP delay-line readout can access both longitudinal energy changes and transverse momentum/deflection signatures. Implementing comparable energy–momentum spectroscopy inside an SEM, however, remains challenging because there is no commercial state-of-the-art diffraction and energy analyser. Achieving this level of measurement in an SEM therefore typically requires custom instrumentation, such as custom designed electrostatic/magnetic lenses and integrated energy- and angle-resolving analysers. A second bottleneck is the effective coherence and phase stability between electrons and the optical field, which is often limited by timing jitter, and drift of the optical phase, especially in pump–probe schemes and long interaction regions. In addition, space-charge effects of the photoexcited electron cloud generated by high-intensity laser fields, as well as laser-induced sample heating, can reduce interaction visibility and distort spectra, particularly in the deep slow-electron regime.\par
In the near term, the most direct experimental progress will be driven by improved in-vacuum photonic setups to enhance electron-light interactions by virtue of Fabry-Pérot photonic cavities, as an example. This includes extending optical programmability to compress electron pulses toward the sub-femtosecond and attosecond regimes; a milestone supported by the quantum limits of free-space laser-based electron energy modulation. Furthermore, implementing reconfigurable longitudinal and transverse wavepacket shaping with light will enable light-based control of electron wavepackets in contrast with massive electron lenses. Treating recoil as an explicit design parameter, alongside improved nanophotonic confinement, will facilitate reproducible access to strong-coupling regimes and support phase-sensitive interferometry with enhanced selectivity.\par
Looking further ahead, several directions are more exploratory and will likely require new levels of quantum control and recoil-aware multi-particle theory. A key longer-term goal is the creation of hybrid flying photon–electron quantum states. Moreover, non-classical or entangled light can imprint correlations onto the electron wavefunction. In parallel, shaped free-electron beams may become active sources of quantum light, generating entangled or non-classical photon states through controlled coupling to nanophotonic structures. Reaching these regimes will require fully quantum electrodynamical and quantum-corrected near-field frameworks that can describe correlated multi-electron and multi-photon processes, and may ultimately open pathways toward preparing and detecting highly nonclassical photon-electron correlated states.\par

In summary, stimulated electron–light interactions have matured from a niche topic into a unifying framework for coherent control, spectroscopy, and quantum information. The continued synthesis of theoretical modeling, ultrafast instrumentation, and nanophotonic design will not only deepen our understanding of light–matter coupling at the quantum level but also pave the way toward a new generation of quantum-coherent electron microscopes capable of writing, reading, and processing information with free electrons, as well as performing interferometry with unprecedented precision.

\newpage

\bibliography{references}   
\bibliographystyle{spiejour}   

\end{document}